\documentclass[11pt]{article}
\usepackage{pstool}
\usepackage[dvipsnames]{pstricks}
\usepackage{epsfig}
\usepackage{pst-grad} 
\usepackage[space]{grffile} 
\usepackage{etoolbox} 
\makeatletter 
\patchcmd\Gread@eps{\@inputcheck#1 }{\@inputcheck"#1"\relax}{}{}
\makeatother

\usepackage{graphicx}
\usepackage{amsmath}
\usepackage{psfrag}
\usepackage{color,soul}
\usepackage{natbib}
\usepackage{siunitx}
\usepackage{float}
\usepackage{overpic}
\usepackage{tikz}
\usepackage{multirow}
\usepackage{bm}
\usepackage{textgreek}
\usepackage{url}
\usepackage{amssymb}
\usepackage{wasysym}
\usepackage{xfrac}
\usepackage{setspace}
\usepackage{gensymb}
\usepackage{textcomp}
\usepackage{mathtools}
\usepackage{arydshln}
\usepackage{subcaption}
\RequirePackage{shellesc}
\usepackage{algorithm}
\usepackage{algpseudocode}
\usepackage{tabularx}
\newenvironment{sloppypar*}
 {\sloppy\ignorespaces}
 {\par}
\usepackage[noblocks]{authblk}
\usepackage[margin=1in]{geometry}
\usepackage[title]{appendix}
\usetikzlibrary{matrix, decorations.pathreplacing, arrows.meta, positioning, calc}
\newcommand\affiliation[1]{\gdef\@affiliation{\let\aff\aff@inst#1}}
\gdef\@affiliation{}
\def\email#1{Email address for correspondence: #1}
\def\aff#1{\ignorespaces\textsuperscript{#1}}
\def\corresp#1{\unskip\thanks{#1}}
\numberwithin{equation}{section}

\renewenvironment{abstract}
{\begin{quote}
\noindent \rule{\linewidth}{.5pt}\par{\bfseries \abstractname.}}
{\medskip\noindent \rule{\linewidth}{.5pt}
\end{quote}
}

  \DeclareTextFontCommand\textsfi{\usefont{OT1}{cmss}{m}{sl}}
  \DeclareMathAlphabet\mathsfi            {OT1}{cmss}{m}{sl}
  \DeclareTextFontCommand\textsfb{\usefont{OT1}{cmss}{bx}{n}}
  \DeclareMathAlphabet\mathsfb            {OT1}{cmss}{bx}{n}
  \DeclareTextFontCommand\textsfbi{\usefont{OT1}{cmss}{m}{sl}}
  \DeclareMathAlphabet\mathsfbi            {OT1}{cmss}{m}{sl}
  \IfFileExists{t1phv.fd}
  {\DeclareTextFontCommand\textsfbi{\usefont{T1}{phv}{b}{it}}
  \DeclareMathAlphabet\mathsfbi            {T1}{phv}{b}{it}
  }{}
  \IfFileExists{ot1phv.fd}
  {\DeclareTextFontCommand\textsfbi{\usefont{OT1}{phv}{b}{it}}
  \DeclareMathAlphabet\mathsfbi            {OT1}{phv}{b}{it}
  }{}
  
\newcommand{\matr}[1]{\mathbf{#1}}
\newcommand{\vectt}[1]{\mathbf{#1}}

\DeclareSymbolFont{matha}{OML}{txmi}{m}{it}

\usepackage{hyperref}

\definecolor{darkblue}{rgb}{0,0,0.80}

\hypersetup{
    colorlinks=true,
    linkcolor={darkblue},
    citecolor={darkblue},
    filecolor=black,
    urlcolor={darkblue},
    plainpages=false,
}

\title{\bf Data-Driven Transient Growth Analysis}

\author[1]{\bf Zhicheng Kai\corresp{\email{kaizz@umich.edu}}}
\author[1]{\bf Peter Frame}
\author[1]{\bf Aaron Towne}

\affil[1]{\normalsize Department of Mechanical Engineering, University of Michigan, Ann Arbor, MI, USA \vspace{-1cm}}  

\date{}
\begin{document}
\maketitle

\begin{abstract}
The transient growth of disturbances made possible by the non-normality of the linearized Navier-Stokes equations plays an important role in bypass transition for many shear flows. Transient growth is typically quantified by the maximum energy growth among all possible initial disturbances, which is given by the largest squared singular value of the matrix exponential of the linearized Navier-Stokes operator. In this paper, we propose a data-driven approach to studying transient growth wherein we calculate optimal initial conditions, the resulting responses, and the corresponding energy growth directly from flow data. Mathematically, this is accomplished by optimizing the growth over linear combinations of input and output data pairs. We also introduce a regularization to mitigate the sensitivity to noisy measurements and unwanted nonlinearity. The data-driven method simplifies and broadens the application of transient growth analysis -- it removes the burden of writing a new code or linearizing an existing one, alleviates the computational expense for large problems, eliminates the challenge of obtaining a well-posed spatial propagator for spatial growth analyses, and enables the direct application of transient growth analysis to experimental data. We validate the data-driven method using a linearized Ginzburg-Landau model problem corrupted by process and measurement noise and obtain good agreement between the data-driven and the standard operator-based results. We then apply the method to study the spatial transient growth of disturbances in a transitional boundary layer using data from the Johns Hopkins Turbulence Database. Our method successfully identifies the optimal output response and provides plausible estimates of the transient spatial energy growth at various spanwise wavenumbers.\\
\end{abstract}


\section{Introduction}
\label{Sec:Intro} 
The classical approach to hydrodynamic stability investigates the long-time asymptotic behavior of infinitesimal disturbances to the base flow of interest. This long-time behavior is determined by the sign of the eigenvalues of the linearized Navier-Stokes (LNS) operator; if there is an eigenvalue with a positive real part, infinitesimal disturbances grow exponentially in the long term; otherwise, they decay. Of course, physical disturbances to the flow are not infinitesimal, and the assumption underlying the relevance of infinitesimal disturbances is that realistic disturbances are initially small enough that nonlinear effects are negligible. This approach to determining stability, usually referred to as modal stability theory, has many successes. For instance, it gives an accurate analytic prediction for the critical Rayleigh number in Rayleigh-Bénard convection \citep{benard1901tourbillons,rayleigh1916lix} and for the critical Taylor number in Taylor-Couette flow \citep{Taylor1923}. However, modal stability theory predicts stability when experiments exhibit turbulence for many wall-bounded shear flows, including many of the most canonical problems in stability theory, such as pipe and Couette flows \citep{tillmark_alfredsson_1992,schmid2002stability, DrazinReid2004}.

\begin{sloppypar}The phenomenon of transient growth \citep{boberg1988onset,butler1992three, Trefethen1993HydrodynamicSW, schmid2002stability} offers a resolution to this issue. The nonnormality of the evolution matrix can cause the disturbances to grow temporally, even in an asymptotically decaying system. This transient growth is usually quantified by the optimal growth $G^{opt}$ and the associated modes representing the optimal initial condition and response. $G^{opt}$ is the maximum growth in the kinetic energy caused by any initial condition. The theory of transient growth has provided physical insight into previously unexplained instabilities in many shear flows in both the temporal~\citep{schmid2002stability} and spatial settings~\citep{reshotko2001transient}.\end{sloppypar}

The standard method for calculating $G^{opt}$ and the associated modes requires access to the LNS operator, but this operator is often difficult to obtain. Deriving the linearized equations and writing a stability code for a new problem can be tedious, as is extracting the linear operator from an existing nonlinear computational fluid dynamics code~\citep{de2012efficient,towne2024resolvent}. If spatial (rather than temporal) growth is of interest, great care must be taken to obtain a stable spatial-evolution operator~\citep{towne2015one,zhu2023recursive}. Once the necessary operator is in hand, computing the optimal growth can be computationally costly, especially for problems that lack homogeneous directions that can be used to decouple individual Fourier modes and reduce the size of the associated operators.

In this paper, we aim to simplify and broaden the application of transient growth analysis by developing an algorithm that approximates $G^{opt}$ using data. The method uses a set of initial- and final-state pairs and assumes that there is an (unknown) linear operator relating the two. With this assumption, the final state resulting from any linear combination of the initial states in the data is the same linear combination of the corresponding final states. The algorithm approximates $G^{opt}$ as the maximum energy ratio of the initial and final state over all such linear combinations. The method also includes a regularization to mitigate the effect of noise. 

The data-driven approach removes the burden of writing a new code or linearizing an existing one. It also substantially alleviates the computational expense -- the proposed method is roughly as expensive as obtaining proper orthogonal decomposition (POD) modes~\citep{Sirovich:1987a}. Finally, it enables the direct application of transient growth analysis to experimental data, which is not possible with the standard operator-based approach.

Our approach differs in a critical way from the work of \cite{dotto2022data}, another recently proposed approach to estimating transient growth from data. \citet{dotto2022data} uses dynamic mode decomposition (DMD) to approximate the one-space-step evolution operator, finds the disturbance that grows the most after one spatial step, and then evolves this disturbance forward in the streamwise direction ($x$) using the same evolution operator at every $x$-location. The optimal transient growth curve, however, is not the energy of any single disturbance as it evolves downstream. Instead, it is the envelope of these energies. The disturbance that grows most rapidly initially is not the disturbance that maximizes the growth at later times. Additionally, the operator that maps a disturbance $\Delta x$ forward in space changes as a function of the $x$-location due to the non-parallel nature of flows investigated with spatial stability. Our approach resolves both issues. 

We first test our method on a linearized Ginzburg-Landau system. We compare the results of our data-driven transient growth algorithm to those obtained using the standard (operator-based) approach and find good agreement even in the presence of significant measurement and process noise. We also use the Ginzburg-Landau system to inform our choice for a parameter that controls the strength of the regularization.

Finally, we apply the method to transitional-boundary-layer data from the Johns Hopkins Turbulence Data Base (JHTDB) \citep{perlman2007data, li2008public,zaki2013streaks,LEE2018142,zhaowu2020}. In this application, we intend to identify spatial energy growth --- the growth of disturbances as they evolve in the streamwise direction. While performing a traditional transient growth analysis is challenging for this problem -- the problem dimension is large, and the spatial LNS operator is not readily available -- the data-driven method scales favorably to large problems and does not require the LNS operator, so it can be readily applied. The $G^{opt}$ results given by the method are in a reasonable range relative to past studies \citep{andersson1999optimal, luchini2000reynolds} on a Blasius boundary layer but are sensitive to a parameter used in the regularization, and are thus uncertain. On the other hand, the output modes we obtain are similar to those of \cite{andersson1999optimal} and \cite{luchini2000reynolds}. In addition, the output modes obtained by the data-driven method at $\beta = 0$ displayed a two-peak structure, which is indicative of modal growth, whereas, at nonzero $\beta$, the modes displayed a one-peak structure indicative of non-modal growth.


The remainder of this paper is organized as follows. We first review the standard (operator-based) transient growth theory in Section~\ref{Sec:second_section} before introducing the data-driven method in the context of temporal stability in Section~\ref{sec:DDTG}. In Section~\ref{sec:Justification}, we use the Ginzburg-Landau equation to verify the accuracy of the data-driven method, assess the impact of various parameters, and determine a recommended value of a key regularization parameter. In Section~\ref{sec:STG}, we apply our method to study the spatial growth of disturbances in a transitional boundary layer, and we conclude the paper in Section~\ref{Sec:Conclusions}.



\section{Transient growth analysis}
\label{Sec:second_section}
The linearized Navier-Stokes equations,
\begin{equation}
    \frac{d}{dt}\vectt{q} = \matr{A}\vectt{q}, \label{eqn:simp_LNS}
\end{equation}
govern the flow disturbance $\vectt{q}=\vectt{q}_{total}-\Bar{\vectt{Q}}\in \mathbb{C}^n$, where $\vectt{q}_{total}$ is the total flow state, $\Bar{\vectt{Q}}$ is the base flow, and ${\bf A} \in \mathbb{C}^{n \times n}$ is the LNS operator.  Here, the state and equations have already been spatially discretized.  
The solution of \eqref{eqn:simp_LNS} is
\begin{equation}
    \vectt{q}_t = e^{\matr{A}t}\vectt{q}_0=\matr{M}_t\vectt{q}_0, \label{eqn:simp_LNS_sol}
\end{equation}
where, $\vectt{q}_t \in \mathbb{C}^n$ is the disturbance $\vectt{q}$ at time $t$, and $\vectt{q}_0$ is the initial disturbance. $\matr{M}_t=e^{\matr{A}t}$ is the time evolution operator that advances the initial condition to time $t$. Equations~\eqref{eqn:simp_LNS} and \eqref{eqn:simp_LNS_sol} lay the foundation for discussing flow stability and energy growth within this paper for the temporal case. In later sections, we also apply the method to the spatial stability case.
The eigenvalues of $\matr{A}$ determine the asymptotic growth rate of the solution of the linearized system~\eqref{eqn:simp_LNS}. If all of the eigenvalues have negative real parts, the norm of the state always decays asymptotically, i.e., $\lim_{t \rightarrow{\infty}}\left\|\vectt{q}_t\right\|=0$. However, if any eigenvalue has a positive real part, the corresponding eigenmode will grow exponentially, and so will the disturbance. Thus, this approach, referred to as modal stability theory, reduces the question of stability to an eigenvalue problem. Modal stability, however, fails to predict the turbulence observed in many shear flows, and the potential for large-scale transient growth offers an explanation.

The key insight is that, though a disturbance to a stable system will decay in the long term, this decay is not required to be monotonic. Instead, as long as the eigenvectors are non-orthogonal, the sum may grow temporarily, and this growth is referred to as transient growth. Indeed, the eigenvectors of the LNS operator in shear flows are often highly non-orthogonal. 

Optimal growth, defined as the maximum energy growth at a given time $t$ across all initial conditions \citep{schmid2002stability}, is often used to quantify transient growth. Mathematically, the optimal growth is expressed as
\begin{equation}
    \label{eqn:gt}
        G^{opt}(t) = \max _{\vectt{q}_0} \frac{\|\vectt{q}_t\|^2}{\left\|\vectt{q}_0\right\|^2}=\max _{\vectt{q}_0} \frac{\left\| \matr{M}_t \vectt{q}_0\right\|^2}{\left\|\vectt{q}_0\right\|^2}.
\end{equation}
The norm, representing the kinetic energy, can be expressed as a weighted $2$-norm,
\begin{equation}
    \left\|\vectt{q}\right\|^2 = \vectt{q}^* \matr{W} \vectt{q},
    \label{eqn:weightedNorm}
\end{equation}
where $\matr{W}$ is the weight matrix and $\left(\cdot\right)^*$ denotes the Hermetian transpose.
Using the Cholesky decomposition $\matr{W} = \matr{L}^*\matr{L}$, the maximized energy growth and its corresponding modes can be obtained from the singular value decomposition \citep{reddy1993energy}
\begin{equation}
    \matr{L} \matr{M}_t \matr{L}^{-1} = \matr{U\Sigma V}^*.
    \label{eqn:growth}
\end{equation}
The optimal growth is given by the largest squared singular value, i.e., $G^{opt}(t)=\sigma_1^2$, and the corresponding input and output modes are the first columns of $\matr{L}^{-1}\matr{U}$ and $\matr{L}^{-1}\matr{V}$, respectively.

\section{Data-driven transient growth analysis}
\label{sec:DDTG}
In this section, we describe a data-driven approach to approximating the transient growth analysis described in the previous section. We formulate the baseline method in Section~\ref{sec:Formulation} before introducing a regularization used to reduce the sensitivity to noisy data and presenting the overall algorithm and its computational cost scaling in Section~\ref{sec:Regularization}.
\subsection{Formulation}
\label{sec:Formulation}
Given data comprised of a set of initial disturbances $\{ \vectt{q}_0^1, \vectt{q}_0^2, ..., \vectt{q}_0^m \}$ and the corresponding evolved disturbances  $\{ \vectt{q}_t^1, \vectt{q}_t^2, ..., \vectt{q}_t^m \}$, where
\begin{equation}
    \vectt{q}_t^k = \matr{M}_t\vectt{q}_0^k, k \in [1,m], \label{eqn:data}
\end{equation}
we form the data matrices
\begin{equation}
\begin{aligned}
    \label{eqn:flowstate}
\matr{Q}_0 = \left[\vectt{q}_0^1, \vectt{q}_0^2, \ldots, \vectt{q}_0^m\right] \in \mathbb{C}^{n \times m},\\
\matr{Q}_t = \left[\vectt{q}_t^1, \vectt{q}_t^2, \ldots, \vectt{q}_t^m\right] \in \mathbb{C}^{n \times m}.
\end{aligned}
\end{equation}

Next, consider a new initial condition constructed as a linear combination of the set of initial disturbances, 
    \begin{equation}
        \vectt{q}_0 = \matr{Q}_0 \vectt{\psi},
        \label{eqn:state_LC_q0}
    \end{equation}
where $\vectt{\psi}\in \mathbb{C}^{m}$ is a vector of expansion coefficients. Since the system is linear, the response at future times to this initial disturbance is obtained from substituting~\eqref {eqn:state_LC_q0} into~\eqref{eqn:data}, 
    \begin{equation}
        \vectt{q}_t = \matr{M}_t\vectt{q}_0 = \matr{M}_t\matr{Q}_0\psi = \matr{Q}_t \vectt{\psi}.
        \label{eqn:state_LC_qt}
    \end{equation}
Substituting~\eqref{eqn:state_LC_q0} and~\eqref{eqn:state_LC_qt} into~\eqref {eqn:gt} gives the (unregularized) data-driven approximation of $G^{opt}$, 
    \begin{align}
        G_{DD}^{opt} =\max _{\vectt{q}_0 \in \mathrm{sp}(\matr{Q}_0)} \frac{\|\vectt{q}_t\|^2  }{\left\|\vectt{q}_0\right\|^2} =\max _{\vectt{\psi}} \frac{\|\matr{Q}_t \vectt{\psi}\|^2  }{\left\|\matr{Q}_0 \vectt{\psi}\right\|^2}.
        \label{eqn:ddtg}
    \end{align}
That is, we seek to maximize the energy growth between initial and evolved states within the span of the data, which is equivalent to maximizing the ratio of the energy of the evolved and initial disturbances over all expansion coefficients. Using the definition of the norm in~\eqref{eqn:weightedNorm}, \eqref{eqn:ddtg} may be rewritten as
\begin{equation}
        G_{DD}^{opt}=\max _{\vectt{\psi}} \frac{\vectt{\psi}^* \matr{Q}_t^* \matr{W} \matr{Q}_t \vectt{\psi}}{\vectt{\psi}^* \matr{Q}_0^* \matr{W} \matr{Q}_0 \vectt{\psi}}.
        \label{eqn:weightednorm}
    \end{equation}

A similar optimization encountered in the context of a data-driven resolvent analysis method of \cite{towne2016advancements} was solved by formulating a cost function and finding its stationary point. Instead, we solve~\eqref{eqn:weightednorm} by transforming it into a Rayleigh quotient.
To this end, we define the Cholesky decomposition
\begin{equation}
   \matr{Q}_0^*\matr{WQ}_0 = \matr{B}^*\matr{B}. \label{eqn:cholesky}
\end{equation}
For now, we assume $\matr{Q}_0$ possesses full column rank, implying that $\matr{Q}_0^* \matr{WQ}_0$ is invertible. This assumption is removed with the regularization presented in the next subsection. Again using $\matr{W}=\matr{L}^* \matr{L}$, ~\eqref{eqn:weightednorm} can be expressed as
\begin{equation}
    G_{DD}^{opt}=\max _{\vectt{v}} \frac{\vectt{v}^* \mathbf{B}^{*-1} \mathbf{Q}_t^*\matr{L}^* \matr{L} \mathbf{Q}_t \mathbf{B}^{-1} \vectt{v}}{\vectt{v}^* \vectt{v}},
\end{equation}
where $\vectt{v}=\matr{B}\vectt{\psi}$ (and $\vectt{\psi}=\matr{B}^{-1}\vectt{v}$). The solution to the Rayleigh quotient is
\begin{equation}
    G_{DD}^{opt} = \sigma_1^2\left(\matr{L}\matr{Q}_t \matr{B}^{-1}\right), \label{eqn:CFM_re}
\end{equation}
and this maximum is achieved when $\vectt{v}=\vectt{v}_1$, which is the first right singular vector of the matrix $\matr{L}\matr{Q}_t \matr{B}^{-1}$.
Finally, the input and output modes can be obtained using the optimal expansion coefficients, $\vectt{\psi}_1=\matr{B}^{-1}\vectt{v}_1$, as
\begin{equation}
\begin{aligned}
    \vectt{q}^{opt}_{0} &= \matr{Q}_0 \vectt{\psi}_1/\left\|\matr{Q}_0 \vectt{\psi}_1\right\|,\\
    \vectt{q}^{opt}_{t} &= \matr{Q}_t \vectt{\psi}_1/\left\|\matr{Q}_t \vectt{\psi}_1\right\|. \label{eqn:IO_2}
\end{aligned}
\end{equation}
In Appendix~\ref{sec:LSAM}, we show that the same result can be obtained using a different formulation inspired by DMD.

\subsection{Regularization}
\label{sec:Regularization}
The method presented in the previous section assumes that the data exactly follow~\eqref{eqn:data}. In realistic use cases, the data may be corrupted with noise or contain the influence of nonlinearity; these effects can be modeled as measurement noise and process noise, respectively. This noise can dominate the small eigenvalues in $\matr{Q}_0^*\matr{WQ}_0$, and, because $\matr{Q}_0^*\matr{WQ}_0$ is in the denominator of the optimization in~\eqref{eqn:ddtg}, these small eigenvalues can dominate the results. Therefore, it is essential to have a regularization method to add robustness to noise.

To this end, we regularize the system by adding a constant to the denominator in~\eqref{eqn:weightednorm},
\begin{equation}
        G_{DD}^{opt}=\max _{\vectt{\psi}} \frac{\vectt{\psi}^* \matr{Q}_t^* \matr{W} \matr{Q}_t \vectt{\psi}}{\vectt{\psi}^* \left(\matr{Q}_0^* \matr{W} \matr{Q}_0 + \gamma\matr{I}\right) \vectt{\psi}},
        \label{eqn:weightednorm_reg}
    \end{equation}
where $\gamma \in \mathbb{R}^+$ serves as the regularization parameter, setting a minimum threshold for the eigenvalues of the matrix in the denominator of the optimization.
Since smaller eigenvalues are often primarily associated with noise, this regularization method efficiently suppresses such unwanted elements without negatively affecting the larger eigenvalues typically associated with coherent structures within the data. The regularization parameter $\gamma$ should be selected to fall between the largest and smallest eigenvalues of $\matr{Q}_0^* \matr{W} \matr{Q}_0$ to ensure minimal interference with essential data. Our strategy for choosing $\gamma$ will be further detailed in Section~\ref{sec:Justification}.
The algorithm then involves performing the Cholesky decomposition
    \begin{equation}
         \left(\matr{Q}_0^* \matr{WQ}_0+\gamma \matr{I}\right)=\matr{B}^*\matr{B}. \label{eqn:reg}
    \end{equation}
Since $\matr{Q}_0^*\matr{WQ}_0 + \gamma I $ is now strictly positive-definite, the Cholesky decomposition is guaranteed to exist, and $\matr{B}$ is guaranteed to be invertible. Compared to the unregularized version,~\eqref{eqn:reg} is the only change in the algorithm.
The method, with the regularization, is given in Algorithm 1.

\begin{algorithm}
\caption{Data-driven transient growth analysis}
\label{alg:2}
\begin{algorithmic}[1]
\State \textbf{Inputs:} $\matr{Q}_0$, $\matr{Q}_t$, $\gamma$
\State $\matr{B} \leftarrow \mathtt{Chol}\left(\matr{Q}_0^*\matr{WQ}_0+\gamma \matr{I}\right)$ 
\State $\left[\matr{U}, \matr{\Sigma} , \matr{V}\right] \leftarrow \mathtt{SVD}\left(\matr{LQ}_t \matr{B}^{-1}\right)$
\State $\psi_1 \leftarrow \matr{B}^{-1} \vectt{v}_1$
\State $G^{opt} \leftarrow \sigma_1^2$ 
\State $\vectt{q}_{0}^{opt} \leftarrow \matr{Q}_0 \psi_1/\left\|\matr{Q}_0 \psi_1\right\|$
\State $\vectt{q}_{t}^{opt} \leftarrow \matr{Q}_t \psi_1/\left\|\matr{Q}_t \psi_1\right\|$
\State \textbf{Outputs:} $G^{opt}, \vectt{q}_{0}^{opt}, \vectt{q}_{t}^{opt}$
\Statex\textbf{Algorithm description.} Inputs: $\matr{Q}_0$, the matrix containing all realizations of initial states; $\matr{Q}_t$, the matrix containing all realizations of states at time $t$; $\gamma$, the regularization parameter. Outputs: $G^{opt}$, the optimal energy growth at time $t$; $\vectt{q}_{0}^{opt}$, $\vectt{q}_{t}^{opt}$ the optimal input and output modes for time $t$.
\end{algorithmic}
\end{algorithm}

Assuming fewer snapshots than degrees of freedom ($m < n$), the most computationally costly step in the algorithm is the SVD in line 2, which scales like $\mathcal{O}(m^2n)$. This is the same scaling as typical algorithms for proper orthogonal decomposition (POD), so application to large datasets should be straightforward. If the algorithm were run with more snapshots than spatial degrees of freedom ($m>n$), which we do not recommend, then the scaling is determined by the Cholesky decomposition in step 1 and inverse in step 3, both of which scale like $\mathcal{O}(m^3)$.

\section{Validation using a linearized Ginzburg-Landau equation}
\label{sec:Justification}
In this section, we apply our method to data generated by the linearized Ginzburg-Landau (GL) equation to assess our approach in a controlled environment and to determine the best choice for the regularization parameter $\gamma$ in the presence of noise. The GL equation offers an ideal testbed for this purpose -- it supports transient growth akin to the linearized Navier-Stokes equations, and the operator-based theory from Section~\ref{Sec:second_section} can be easily applied for comparison against results from the data-driven method from Section~\ref{sec:DDTG}. 

\subsection{Ginzburg-Landau equation and data matrix construction}
\label{sec:GL}
\begin{figure}[t]
    \centering
    \input{figure/plt2_GLresult}
    \includegraphics{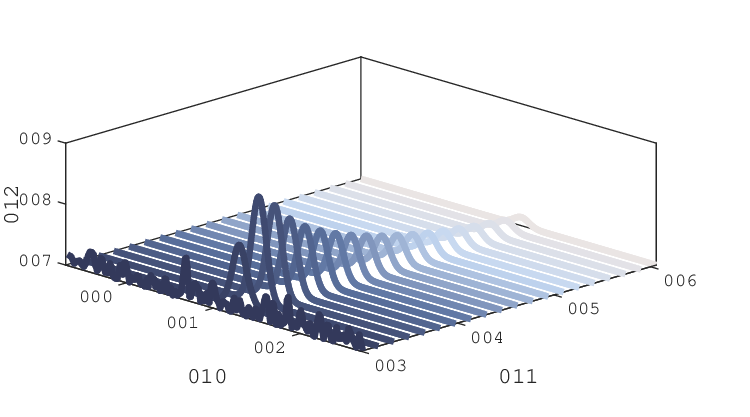}
    \caption{One trajectory of the GL equation with $\mu_0 = 0.38$ for a random initial condition. The trajectory displays transient energy growth before an exponential decay.}
    \label{fig:GLresult}
\end{figure}

The GL equation is
\begin{equation}
    \begin{aligned}
    &\frac{\partial q}{\partial t}=\mathcal{A}q(x)=\left(-\nu \frac{\partial}{\partial x}+\gamma_g \frac{\partial^2}{\partial x^2}+\mu(x)\right) q,
    \end{aligned}
\end{equation}
where $\nu$ and $\gamma_g$ are the flow convection and dissipation coefficients, respectively, and $\mu(x) = \mu_0-c_u^2+\mu_2 \frac{x^2}{2}$ produces local exponential growth or decay depending on the sign of $\mu(x)$.
Following \cite{bagheri2009input}, we set $\nu=2+0.4i, \gamma_g = 1-i, \mu_0 = 0.38, c_u = 0.2,$ and $\mu_2 = -0.01$.
The equation is spatially discretized using $n = 220$ Hermite polynomials and integrated in time using a fourth-order Runge-Kutta scheme. Figure~\ref{fig:GLresult} shows one trajectory of the Ginzburg-Landau system. As time progresses, a transient increase in energy is visible before the eventual asymptotic decay.

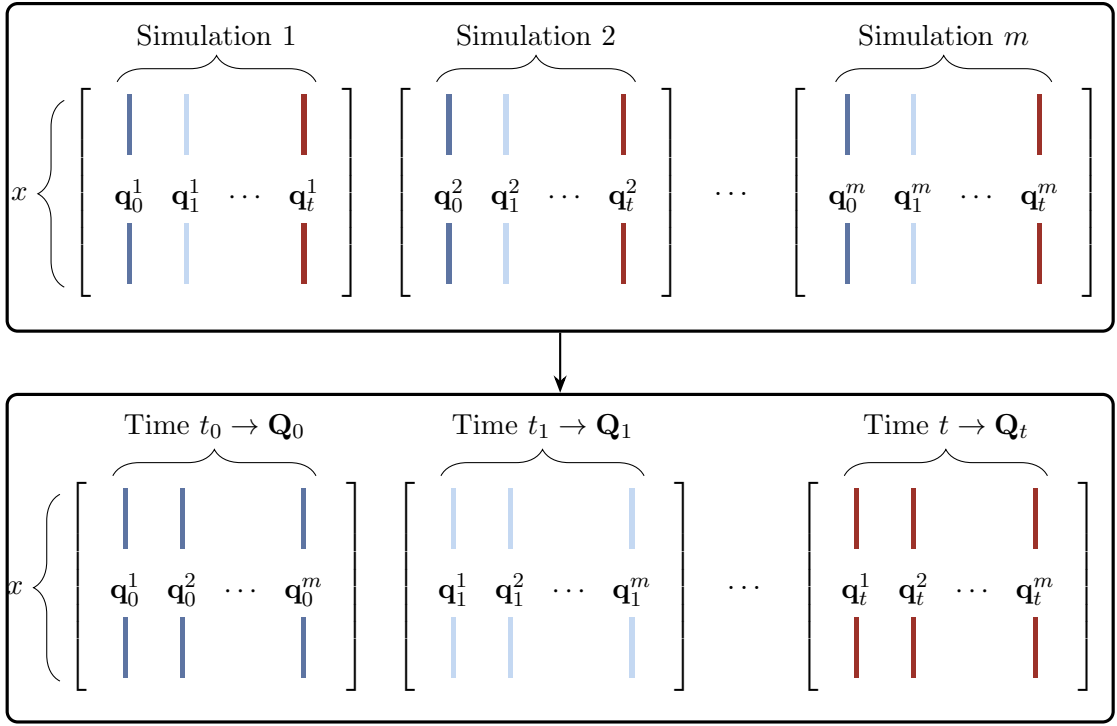
\begin{figure}[t]
    \centering
    \usetikzlibrary{fit}

\begin{tikzpicture}[scale=1, transform shape=false, node/.style={font=\normalsize}]

\matrix[matrix of math nodes, nodes in empty cells, row sep=0.1cm, column sep=0.1cm,nodes={text height=1em, text depth=0.3em}, left delimiter={[}, right delimiter={]}] (m1) {
     \color[rgb]{0.364705882352941,0.454901960784314,0.635294117647059}\rule[-2mm]{1.5pt}{8mm} & \color[rgb]{0.768627450980392,0.847058823529412,0.949019607843137}\rule[-2mm]{1.5pt}{8mm}& & \color[rgb]{0.611764705882353,0.192156862745098,0.160784313725490}\rule[-2mm]{1.5pt}{8mm} \\
    \vectt{q}_0^1 & \vectt{q}_1^1 & \cdots & \vectt{q}_t^1 \\
    \color[rgb]{0.364705882352941,0.454901960784314,0.635294117647059}\rule[-2mm]{1.5pt}{8mm} & \color[rgb]{0.768627450980392,0.847058823529412,0.949019607843137}\rule[-2mm]{1.5pt}{8mm}& & \color[rgb]{0.611764705882353,0.192156862745098,0.160784313725490}\rule[-2mm]{1.5pt}{8mm} \\
};

\matrix[matrix of math nodes, nodes in empty cells, row sep=0.1cm, column sep=0.1cm,nodes={text height=1em, text depth=0.3em}, left delimiter={[}, right delimiter={]}, right=1cm of m1] (m2) {
     \color[rgb]{0.364705882352941,0.454901960784314,0.635294117647059}\rule[-2mm]{1.5pt}{8mm} & \color[rgb]{0.768627450980392,0.847058823529412,0.949019607843137}\rule[-2mm]{1.5pt}{8mm}& & \color[rgb]{0.611764705882353,0.192156862745098,0.160784313725490}\rule[-2mm]{1.5pt}{8mm} \\
    \vectt{q}_0^2 & \vectt{q}_1^2 & \cdots & \vectt{q}_t^2 \\
     \color[rgb]{0.364705882352941,0.454901960784314,0.635294117647059}\rule[-2mm]{1.5pt}{8mm} & \color[rgb]{0.768627450980392,0.847058823529412,0.949019607843137}\rule[-2mm]{1.5pt}{8mm}& & \color[rgb]{0.611764705882353,0.192156862745098,0.160784313725490}\rule[-2mm]{1.5pt}{8mm} \\
};

\matrix[matrix of math nodes, nodes in empty cells, row sep=0.1cm, column sep=0.1cm,nodes={text height=1em, text depth=0.3em}, left delimiter={[}, right delimiter={]}, right=2cm of m2] (m3) {
     \color[rgb]{0.364705882352941,0.454901960784314,0.635294117647059}\rule[-2mm]{1.5pt}{8mm} & \color[rgb]{0.768627450980392,0.847058823529412,0.949019607843137}\rule[-2mm]{1.5pt}{8mm}& & \color[rgb]{0.611764705882353,0.192156862745098,0.160784313725490}\rule[-2mm]{1.5pt}{8mm} \\
    \vectt{q}_0^m & \vectt{q}_1^m & \cdots & \vectt{q}_t^m \\
     \color[rgb]{0.364705882352941,0.454901960784314,0.635294117647059}\rule[-2mm]{1.5pt}{8mm} & \color[rgb]{0.768627450980392,0.847058823529412,0.949019607843137}\rule[-2mm]{1.5pt}{8mm}& & \color[rgb]{0.611764705882353,0.192156862745098,0.160784313725490}\rule[-2mm]{1.5pt}{8mm} \\
};

\matrix[matrix of math nodes, nodes in empty cells, row sep=0.1cm, column sep=0.1cm,nodes={text height=1em, text depth=0.3em}, left delimiter={[}, right delimiter={]}, below=2.5cm of m1] (m4) {
     \color[rgb]{0.364705882352941,0.454901960784314,0.635294117647059}\rule[-2mm]{1.5pt}{8mm} & \color[rgb]{0.364705882352941,0.454901960784314,0.635294117647059}\rule[-2mm]{1.5pt}{8mm}& & \color[rgb]{0.364705882352941,0.454901960784314,0.635294117647059}\rule[-2mm]{1.5pt}{8mm} \\
    \vectt{q}_0^1 & \vectt{q}_0^2 & \cdots & \vectt{q}_0^m \\
     \color[rgb]{0.364705882352941,0.454901960784314,0.635294117647059}\rule[-2mm]{1.5pt}{8mm} & \color[rgb]{0.364705882352941,0.454901960784314,0.635294117647059}\rule[-2mm]{1.5pt}{8mm}& & \color[rgb]{0.364705882352941,0.454901960784314,0.635294117647059}\rule[-2mm]{1.5pt}{8mm} \\
};
\matrix[matrix of math nodes, nodes in empty cells, row sep=0.1cm, column sep=0.1cm,nodes={text height=1em, text depth=0.3em}, left delimiter={[}, right delimiter={]}, right=1cm of m4] (m5) {
     \color[rgb]{0.768627450980392,0.847058823529412,0.949019607843137}\rule[-2mm]{1.5pt}{8mm} & \color[rgb]{0.768627450980392,0.847058823529412,0.949019607843137}\rule[-2mm]{1.5pt}{8mm}& & \color[rgb]{0.768627450980392,0.847058823529412,0.949019607843137}\rule[-2mm]{1.5pt}{8mm} \\
    \vectt{q}_1^1 & \vectt{q}_1^2 & \cdots & \vectt{q}_1^m \\
     \color[rgb]{0.768627450980392,0.847058823529412,0.949019607843137}\rule[-2mm]{1.5pt}{8mm} & \color[rgb]{0.768627450980392,0.847058823529412,0.949019607843137}\rule[-2mm]{1.5pt}{8mm}& & \color[rgb]{0.768627450980392,0.847058823529412,0.949019607843137}\rule[-2mm]{1.5pt}{8mm} \\
};
\matrix[matrix of math nodes, nodes in empty cells, row sep=0.1cm, column sep=0.1cm,nodes={text height=1em, text depth=0.3em}, left delimiter={[}, right delimiter={]}, right=2cm of m5] (m6) {
     \color[rgb]{0.611764705882353,0.192156862745098,0.160784313725490}\rule[-2mm]{1.5pt}{8mm} & \color[rgb]{0.611764705882353,0.192156862745098,0.160784313725490}\rule[-2mm]{1.5pt}{8mm}& & \color[rgb]{0.611764705882353,0.192156862745098,0.160784313725490}\rule[-2mm]{1.5pt}{8mm} \\
    \vectt{q}_t^1 & \vectt{q}_t^2 & \cdots & \vectt{q}_t^m \\
     \color[rgb]{0.611764705882353,0.192156862745098,0.160784313725490}\rule[-2mm]{1.5pt}{8mm} & \color[rgb]{0.611764705882353,0.192156862745098,0.160784313725490}\rule[-2mm]{1.5pt}{8mm}& & \color[rgb]{0.611764705882353,0.192156862745098,0.160784313725490}\rule[-2mm]{1.5pt}{8mm} \\
};
\node[draw=black, very thick, rounded corners, inner sep=23pt, yshift=10pt, xshift=-10pt, fit=(m1) (m2) (m3)] (box) {};
\node[draw=black, very thick, rounded corners, inner sep=23pt, yshift=10pt, xshift=-10pt, below=5.2cm of box, fit=(m1) (m2) (m3)] (box2) {};

    
\draw[decorate,decoration={brace,amplitude=10pt}] 
    ([xshift=-20pt]m1-3-1.south west) -- ([xshift=-20pt]m1-1-1.north west) 
    node[midway,left=10pt] {$x$};

\draw[decorate,decoration={brace,amplitude=10pt}] 
    ([xshift=-20pt]m4-3-1.south west) -- ([xshift=-20pt]m4-1-1.north west) 
    node[midway,left=10pt] {$x$};

\draw[decorate,decoration={brace,amplitude=10pt}] 
    ([yshift=7pt]m1-1-1.north west) -- ([yshift=7pt]m1-1-4.north east) 
    node[midway,above=10pt] {Simulation 1};

\draw[decorate,decoration={brace,amplitude=10pt}] 
    ([yshift=7pt]m2-1-1.north west) -- ([yshift=7pt]m2-1-4.north east) 
    node[midway,above=10pt] {Simulation 2};

\draw[decorate,decoration={brace,amplitude=10pt}] 
    ([yshift=7pt]m3-1-1.north west) -- ([yshift=7pt]m3-1-4.north east) 
    node[midway,above=10pt] {Simulation $m$};

\node at ($(m2.east)!0.5!(m3.west)$) {\dots};
\node at ($(m5.east)!0.5!(m6.west)$) {\dots};

\draw[decorate,decoration={brace,amplitude=10pt}] 
    ([yshift=7pt]m4-1-1.north west) -- ([yshift=7pt]m4-1-4.north east) 
    node[midway,above=10pt] {Time $t_0 \to \matr{Q}_0$};

\draw[decorate,decoration={brace,amplitude=10pt}] 
    ([yshift=7pt]m5-1-1.north west) -- ([yshift=7pt]m5-1-4.north east) 
    node[midway,above=10pt] {Time $t_1 \to \matr{Q}_1$};

\draw[decorate,decoration={brace,amplitude=10pt}] 
    ([yshift=7pt]m6-1-1.north west) -- ([yshift=7pt]m6-1-4.north east) 
    node[midway,above=10pt] {Time $t \to \matr{Q}_t$};


\draw [-Stealth, thick] (box.south) -- (box2.north);
\end{tikzpicture}
    \caption{Constructing the data matrices. The matrices in the first row represent different trajectories of the solution obtained from different initial conditions, with each column color-coded according to its position in time. These columns are sorted by time in the second row, forming $\matr{Q}_0$ and $\matr{Q}_t \in \mathbb{C}^{n \times m}$.}
    \label{fig:GL_Ds}
\end{figure}

To form the data matrices in~\eqref{eqn:flowstate}, we collect $m$ realizations of the data pairs $[\vectt{q}_{0}^i, \vectt{q}_{t}^i]$, with $t \in \left[0, T\right]$, by computing $m$ trajectories of duration $T$, each starting from a different randomly generated initial condition. We choose initial conditions drawn from a multivariate Gaussian distribution with zero mean and spatial correlation of the form \citep{frame2024beyond}
\begin{equation}
\mathbb{E}[q_0(x_1) q_0^*(x_2)] = \exp \left[-\frac{(x_1-x_2)^2}{\lambda^2} \right] \text{,}
\label{eqn:gaus}
\end{equation}
with a correlation length of $\lambda = 2$.  Each initial condition comprises one column of $\matr{Q}_0$, and we then evolve each initial condition to generate $\matr{Q}_t$ at subsequent times. Figure~\ref{fig:GL_Ds} illustrates this process. The number of realizations should be large enough such that the data matrices are close to spanning the desired optimal input and output modes; this is assessed in Section~\ref{sec:GL_result}.



To assess the resilience of the regularized method to noise, we introduce both measurement and process noise. After spatial discretization, the noisy system is
\begin{align}
    \label{eqn:GL_noised}
    \frac{d\vectt{q}}{dt}=\matr{A}\vectt{q}+\matr{W}&,\\
    \Tilde{\vectt{q}} = \vectt{q}+\matr{N}&,
\end{align}
where $\vectt{W}$ and $\vectt{N}$ represent the process and  measurement noise, respectively, and  $\Tilde{\vectt{q}}$ denotes the noisy measurement of the state. Both types of noise are spatially Gaussian, as in \eqref{eqn:gaus}. The measurement noise is white-in-time, while the process noise is band-limited white noise (between frequencies $\omega = 0.001$ and $\omega = 0.5$) to avoid numerical issues when integrating \eqref{eqn:GL_noised}.

\subsection{Error metrics}
\label{sec:GL_error}
We evaluate the data-driven method by comparing its predicted growth and input and output modes against those computed using the standard operator-based approach, which is straightforward to apply to the Ginzburg-Landau problem. We quantify the error using two metrics.

First, the peak growth error is defined as
    \begin{equation}
        \label{eqn:diffrel}
        \epsilon_G = \frac{\left|G_{\max, DD}^{opt}-G_{\max}^{opt}\right|}{G_{\max}^{opt}},
    \end{equation}
where $G_{max}^{opt}$ indicates the largest optimal growth over all time, and the subscript $DD$ refers to the data-driven results. This metric quantifies the relative discrepancy between the peak transient growth values predicted by the data-driven and operator-based methods.

Second, the error in the input and output modes is defined, respectively, as
    \begin{equation}
    \begin{aligned}
        \epsilon_0^{opt}= 1-\left|\left<\vectt{q}_{DD, i}, \vectt{q}_i\right>\right|,\\
        \epsilon_t^{opt}= 1-\left|\left<\vectt{q}_{DD, o}, \vectt{q}_o\right>\right|,
    \end{aligned}\label{eqn:epsi}
    \end{equation}
where $\vectt{q}$ is the state vector. The scalars $\epsilon_0^{opt}$ and $\epsilon_t^{opt}$ are $1$ if the modes are orthogonal and zero if they are the same up to an arbitrary complex phase.

\subsection{Results}
\label{sec:GL_result}
In this subsection, we test our data-driven method using the Ginzburg-Landau equation. First, we look at the effect of the number of realizations, then evaluate the impact of the regularization parameter, and propose a standard for selecting it.

The number of realizations $m$ represents the number of independent state vectors evolving over time. The set of realizations should be large enough to accurately span the desired optimal input and output modes. At the same time, the number of realizations may be constrained in practice by the dataset in hand. Moreover, while a larger number of realizations may always appear desirable, we will see that using more data than necessary increases the sensitivity of the method to noise. Overall, these considerations highlight the need for a technique that is effective across a range of $m$ values.

\begin{figure}[t]
        \centering
        \input{figure/plt_GL_0noise_fourm}
        \includegraphics{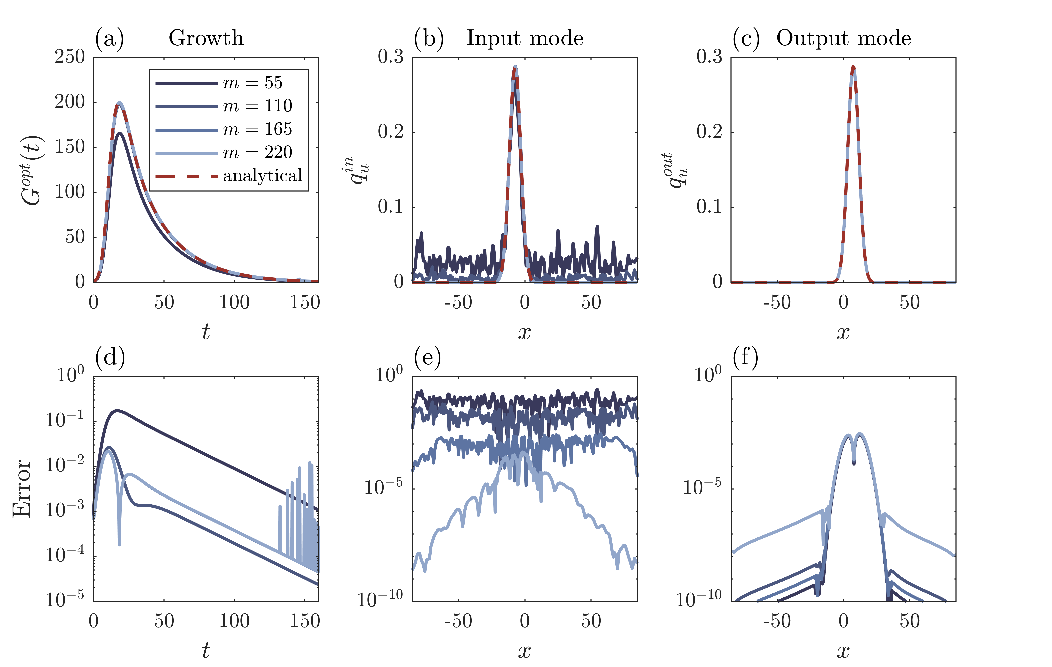}
        \caption{Comparison between the (unregularized) data-driven and operator-based transient growth results for the Ginzburg-Landau system without noise: (a) optimal growth; (b) optimal input mode; (c) optimal output mode; (d) growth error; (e) input mode error; (f) output mode error.   Error is computed as the difference between the data-driven and operator-based results.}
        \label{fig:GL_0noise}
    \end{figure}

We first evaluate the impact of $m$ in the absence of noise. Typically, one should avoid making $m$ close to or larger than $n$, as doing so leads to redundancy and computational inefficiency. However, for the sake of exploration, we experimented with $m = [55, 110, 165, 220]$, which ranges from $n/4$ to $n$. Figure~\ref{fig:GL_0noise} shows the optimal transient growth and its optimal modes for these four values of $m$ using the unregularized method, as well as the operator-based, i.e., ground truth, results.
The transient growth plot indicates that the estimation improves as $m$ increases, with smaller $m$ values yielding lower growth values. It is reasonable that with a small $m$ and zero noise, the optimal growth is lower since fewer combinations of vectors are available to maximize over, i.e., the data do not fully span the desired operator-based optimal modes.
The output modes are more accurately captured than the input modes. This is a consequence of the growth of the linear operator; the part of each realization of the initial condition that projects onto the optimal input mode is amplified by a factor of $G^{opt}$, so the mode we are trying to represent is more prevalent in the data. A close match with the input mode is achieved when $n=m$, while the results for the output mode closely align with the analytical outcome for all values of $m$.

Next, we introduce moderate noise levels ($\mathbb{E}\left[\matr{\left\|W\right\|}\right]=\mathbb{E}\left[\matr{\left\|N\right\|}\right]=0.03$). Since our initial conditions have expected energy of $1$, the expectation of noise can be seen as relative to that of the initial condition.
We use a regularization factor of $\gamma = \gamma_0 \lambda_1\left(\matr{Q}_0^*\matr{WQ}_0\right)$ where $\gamma_0 = \left[0, 0.01, 0.02, 0.04\right]$. In other words, we choose the regularization parameter to be proportional to the largest eigenvalue of $\matr{Q}_0^*\matr{WQ}_0$ with a constant of proportionality of $\gamma_0$.

    \begin{figure}[t]
        \centering
        \input{figure/plt_GL_m}
        \includegraphics{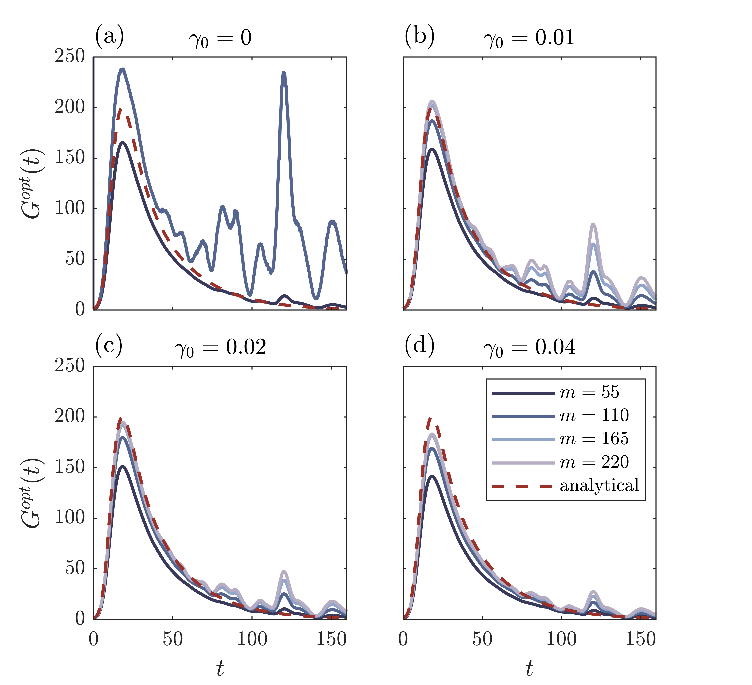}
        \caption{Data-driven approximations of transient growth for the Ginzburg-Landau system with 3\% noise for different numbers of realizations $m$ and choices of the regularization parameter $\gamma_0$.}\label{fig:plt_GL_m}
    \end{figure}
    Figure~\ref{fig:plt_GL_m} compares the optimal operator-based and data-driven transient growth for several values of $\gamma_0$ and $m$. Each subfigure shows how the growth changes with $m$ under a particular value of the regularization parameter $\gamma_0$.
    We observe that larger $m$ increases the potential of the growth regardless of the value of $\gamma_0$, which is due to the availability of more combinations of vectors to maximize over. However, a larger $m$ also increases the effect of noise. As shown in Figure~\ref{fig:plt_GL_m}, although larger $\gamma_0$ helps to suppress the noise, we can still observe that as $m$ increases, the growth results display noisy structures, especially when the growth is small (at $t>100$).
   
    \begin{figure}[t]
        \centering
        \input{figure/plt_GL_m_err}
        \includegraphics{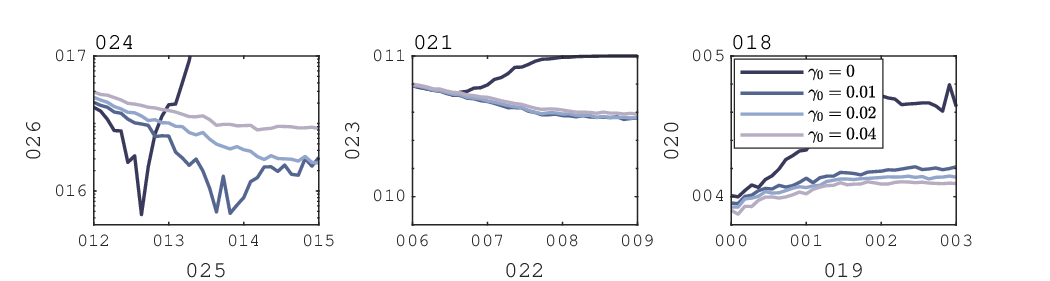}
        \caption{Growth and mode errors for the Ginzburg-Landau system corrupted with 3\% measurement and process noise as a function of the number of realizations $m$ for several values of the regularization parameter $\gamma_0$: (a) peak growth error $\epsilon_G$; (b) input mode error $\epsilon_{0}^{opt}$; and (c) output mode error $\epsilon_{t}^{opt}$.}\label{fig:plt_GL_m_err}
    \end{figure}

Figure~\ref{fig:plt_GL_m_err} quantifies the error in the growth, input modes, and output modes for the Ginzburg-Landau equation corrupted by measurement and process noise using the metrics $\epsilon_G$, $\epsilon_0^{opt}$, and $\epsilon_t^{opt}$, respectively, as defined in \eqref{eqn:diffrel} and \eqref{eqn:epsi}.  Unlike the noise-free case, the unregularized method ($\gamma_0 = 0$) does not converge with increasing $m$; instead, the error temporarily drops before diverging toward high error levels as more realizations are added. In contrast, the regularized method ($\gamma_0 > 0$) is relatively insensitive to the number of realizations; the growth and input mode errors decrease slightly, and the output mode error increases slightly with increasing $m$, but the absolute error levels are reasonable in all cases. Additionally, the results are insensitive to the size of the regularization parameter within the range we consider. As in the noise-free case, the error is especially low for the output modes. Overall, these results highlight the critical need for regularization to ensure reliable results when the data is corrupted by noise.

As previously mentioned, we select the regularization parameter $\gamma$ based on the maximum eigenvalue of $\matr{Q}_0^*\matr{WQ}_0$, such that $\gamma = \gamma_0 \lambda_1\left(\matr{Q}_0^*\matr{WQ}_0\right)$, so that a given value of $\gamma_0$ has the same meaning regardless of the number of realizations or the energy of the initial disturbances. The sensitivity of the method to noise stems from the significant disparity between the maximum and minimum eigenvalues of $\matr{Q}_0^* \matr{WQ}_0$. The smallest eigenvalues can be easily affected by noise, and, because the algorithm involves inverting the Cholesky factor of $\matr{Q}_0^*\matr{WQ}_0$, these eigenvalues can have a significant impact on the outputs of the algorithm.

\begin{figure}[t]
    \centering
    \input{figure/plt_mu038mxgammaIO}
    \includegraphics{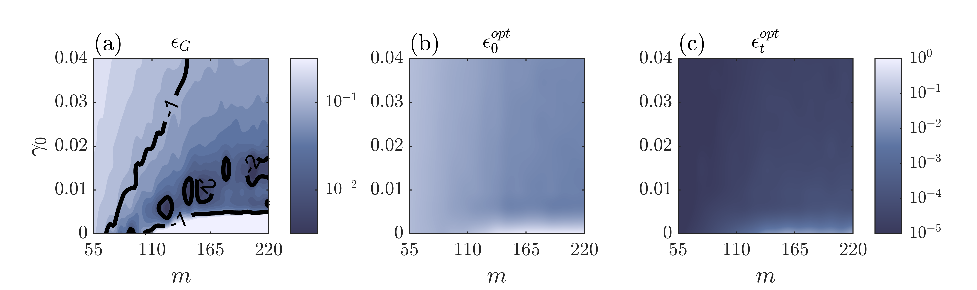}
    \caption{Growth and mode errors for the Ginzburg-Landau system corrupted with 3\% measurement and process noise as a function of the number of realizations $m$ and the regularization parameter $\gamma_0$: (a) peak growth error $\epsilon_G$; (b) input mode error $\epsilon_{0}^{opt}$; and (c) output mode error $\epsilon_{t}^{opt}$. In (a), $\epsilon_G=10^{-2}$ and $\epsilon_G=10^{-1}$ are marked in bold. }\label{fig:IO_iterMgamma}
\end{figure}
We show the effect of $m$ and $\gamma_0$ on our estimates of peak growth and optimal modes in Figure~\ref{fig:IO_iterMgamma}. Figure~\ref{fig:IO_iterMgamma}(a) shows how the peak growth error $\epsilon_G$ varies with $m$ and $\gamma_0$.
We can observe that for $m>140$ and $0.005<\gamma_0<0.02$, the error becomes mostly independent of $m$.
However, $\epsilon_0^{opt}$ and $\epsilon_t^{opt}$ exhibit a different dependencies on $\gamma$ and $m$.
As shown in Figure~\ref{fig:IO_iterMgamma}(b) and (c), the errors for both input and output modes are nearly independent of $\gamma_0$ and $m$, except when $\gamma_0$ is very low (for $\epsilon_t^{opt}$) or when either $m$ or $\gamma_0$ is very low (for $\epsilon_0^{opt}$). As previously noted, the input mode is not amplified by the singular values, which explains why $\epsilon_0^{opt}$ tends to be higher.
In contrast, the output mode can be estimated accurately with fewer realizations, and larger $m$ values negatively affect the results.
These observations regarding $\epsilon_0^{opt}$ and $\epsilon_t^{opt}$ also indicate that there is a flexible range for selecting $\gamma_0$, as long as it is above a specific threshold.

\begin{figure}[t]
        \centering
        \input{figure/plt_IOiterNoise}
        \includegraphics[width = 6in]{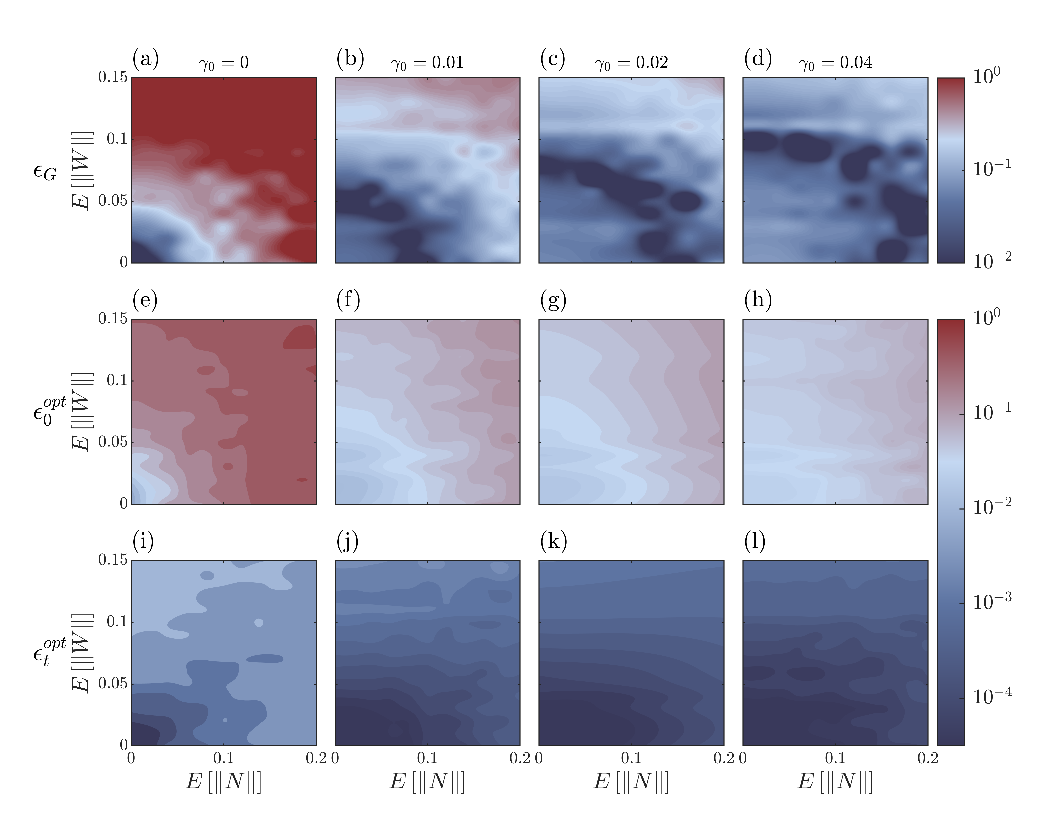}
        \caption{Growth and mode errors for the Ginzburg-Landau system as a function of the measurement and process noise levels: (top row) peak growth error $\epsilon_G$; (middle row) input mode error $\epsilon_{0}^{opt}$; and (bottom row) output mode error $\epsilon_{t}^{opt}$. Columns from left to right: $\gamma_0 = [0, 0.01, 0.02, 0.04]$. All cases use $m=110$ realizations.}\label{fig:changenoise}
    \end{figure}

Next, we select a few typical $\gamma_0$ values to evaluate their effectiveness across different noise levels. Specifically, in Figure~\ref{fig:changenoise}, we show each error metric as a function of the mean amplitude of the measurement and process noise for $\gamma_0=[0,0.01,0.02,0.04]$. The first column of the figure shows that an unregularized approach ($\gamma_0=0$) closely matches analytical predictions in the absence of noise. However, unsurprisingly, its effectiveness decreases as the noise levels increase. The remaining three columns of Figures~\ref{fig:changenoise} show that all three regularized cases ($\gamma_{0}>0$) provide accurate estimates of the maximum transient growth and the optimal modes in the presence of noise. This analysis confirms that, for the GL equation, our method delivers reasonable results within a flexible parameter range.

\section{Spatial transient growth in a transitional boundary layer}
\label{sec:STG}
The previous section demonstrates the robustness of our method on the GL equation and provides guidance for selecting the regularization parameter $\gamma_0$. In this section, we use this method to study spatial transient growth in a boundary layer. The relevant growth in boundary layers is spatial rather than temporal, i.e., disturbances grow as they advect downstream in the boundary layer, and this process is statistically stationary in time. Thus far, we have discussed temporal transient growth, so we first review spatial transient growth theory in Section~\ref{Sec:stg_intro}. We then describe the boundary layer data to which we apply our method in Section~\ref{sec:BLdata} and describe the data processing required to construct the data matrices in the spatial case in Section~\ref{sec:BL_pre}. Finally, in Section~\ref{sec:BL_re}, we report our results, i.e., the growth and modes for the boundary layer.
\subsection{Spatial transient growth} \label{Sec:stg_intro}
The state vector in the spatial case can be decomposed as
\begin{equation}
    \boldsymbol{q}(x, y, z, t)=\int_{-\infty}^\infty \int_{-\infty}^\infty \hat{\boldsymbol{q}}(x, y, \beta, \omega) \exp [i(\beta z-\omega t)]d\beta d\omega, \label{eqn:stateFT}
\end{equation}
where $\hat{\boldsymbol{q}}(x,y,\beta,\omega)$ is the velocity field Fourier transformed in $z$ and $t$, $\beta$ is the spanwise wavenumber and $\omega$ is the frequency. Upon discretization in the $y$-direction, the state at a certain $x$, $\beta$, and $\omega$ is $\hat{\vectt{q}}\left(x, \beta, \omega \right)$. The transformation is standard and is useful because each $\left(\beta, \omega\right)$ pair evolves independently, i.e., $\hat{\vectt{q}}\left(x, \beta, \omega \right)$ is a linear function of $\hat{\vectt{q}}\left(0, \beta, \omega \right)$ alone. This is useful for two reasons. First, it reduces computation time, as analyzing the data at a specific $\left(\beta, \omega\right)$ pair reduces the original four-dimensional data matrix to a two-dimensional data matrix. Second, it enables examination of flow properties at various frequencies and wavenumbers and facilitates comparisons with results from modal analyses.

Analogous to the temporal evolution equation~\eqref{eqn:simp_LNS}, the spatial evolution of the state for each $\left(\beta, \omega\right)$ can be written as
\begin{equation} \label{eq:spatial_tg_evolution}
    \frac{d}{dx} \vectt{q} = \matr{A}(x) \vectt{q}.
\end{equation}
Here, and in what follows, we have dropped the $( \ \hat{\cdot} \ )$ and suppressed the arguments for notational convenience, i.e., $\vectt{q}$ is the $y$-discretized state at a certain $x$, $\beta$, and $\omega$. 
An important difference between the temporal evolution equation in~\eqref{eqn:simp_LNS} and the spatial evolution equation in ~\eqref{eq:spatial_tg_evolution} is that, in the latter, the linear operator is a function of the direction of evolution, i.e., ${\matr{A}}$ is not a function of $t$ in the temporal case but is a function of $x$ in the spatial case. Nevertheless, because the equation is linear, an input-output map exists of the form
\begin{equation}
    \vectt{q}_x = \matr{M}_x \vectt{q}_0. \label{eqn:BL_state}
\end{equation}
Here, $ \vectt{q}_x$ is the state at downstream position $x$, $\vectt{q}_0$ is the state at an initial upstream position, and $\matr{M}_x$ is the linear operator that relates the two. This equation is now entirely analogous to the temporal case in~\eqref{eqn:simp_LNS_sol}, and we may apply the proposed algorithm. While the spatial propagator is cumbersome to obtain from first principles~\citep{towne2015one,towne2022efficient}, making an operator-based approach difficult, our data-driven approach may be applied with little additional difficulty relative to the temporal case.



\subsection{Boundary layer data}
\label{sec:BLdata}
\begin{figure}[t]
    \centering
    \input{figure/plt_cf}
    \includegraphics[width=6in]{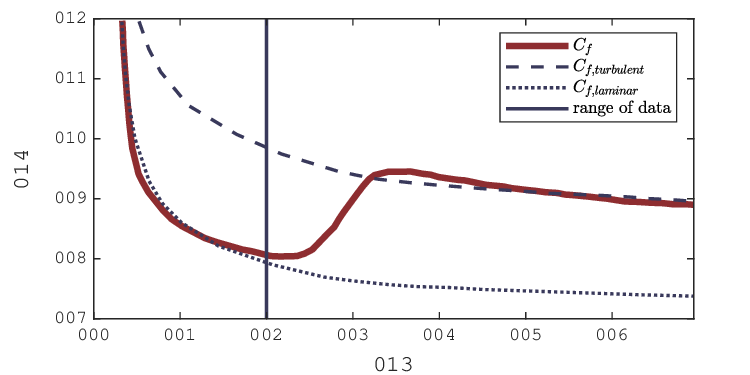}
    \caption{Skin-friction coefficient computed from the data compared to laminar and turbulent values. The flow begins to transition near $Re_x = 2\times 10^5$. For this study, we use data to the left of the vertical line. }
    \label{fig:BL_cf}
\end{figure}

\begin{sloppypar}The boundary layer data is obtained from the Johns Hopkins Turbulence Database (JHTDB)~\citep{perlman2007data, li2008public, zaki2013streaks,LEE2018142,zhaowu2020}. The Reynolds number based on the half-plate thickness $L_p$ is $Re_{L_p}=800$. The boundary layer undergoes bypass transition beginning at a distance $x \approx 350L_p$ downstream of the leading edge of the plate \citep{LEE2018142}, as indicated by the skin friction coefficient shown in Figure~\ref{fig:BL_cf}.\end{sloppypar}

All data in the database is non-dimensionalized using the plate half-thickness $L_p$ and the origin defined at the leading edge, but these are not the natural choices for stability analyses. In preparation for using our algorithm and comparing our results to those in the stability literature, we transform the data as follows.
First, we shift the $x$-coordinate to the best match between the mean streamwise velocity calculated from the data and the Blasius solution \citep{kai2024data}. The shifted $x$-coordinate is used to define the Reynolds number,
\begin{equation}
    Re_x=\frac{U x}{\nu}=Re_{L_p} \frac{x}{L_p}. \label{fig:Rex}
\end{equation}
For consistency with the literature, we non-dimensionalize the wall-normal and spanwise directions according to the boundary layer displacement thickness from a Blasius profile \citep{andersson1999optimal, luchini2000reynolds}.
Specifically, $y$ and $z$ are non-dimensionlized as $y/\delta$ and $z/\delta$, respectively, where $\delta\left(x\right) = \sqrt{\nu x / U_{\infty}}$ is the $x$-dependent transverse length scale. The parameter $\delta$ is related to the Blasius boundary layer displacement thickness as $\delta^* \approx 1.72\delta$. In the later discussion, $\delta_{x_0}$ is the boundary layer thickness at the first $x$-location in the data, while $\delta$ refers to the local thickness at the $x$ location of interest.
The dimensionless spanwise wavenumber is then 
\begin{equation}
    \beta \delta = \left(\beta L_p\right) \left(Re_{L_p}\right)^{-1/2} \sqrt{\frac{x}{L_p}},
\end{equation}
where $\beta_{L_p}$ is the spanwise wavenumber implied by the original non-dimensionalization of the database. Finally, the frequency is non-dimensionalized as $\omega x_0/U_\infty$.

Transient growth analysis, including our data-driven method, assumes linear dynamics. Accordingly, we confine our analysis to the laminar region of the flow where the nonlinearity is small, i.e., we focus on linear growth of disturbances that preceded transition. This range can be estimated by looking at the skin friction coefficient given by \cite{LEE2018142}, as shown in Figure~\ref{fig:BL_cf}. We used data up to $Re_x=2\times 10^5$, i.e., to the left of the vertical line in the figure.

\subsection{Data matrix construction for the boundary layer data}
\label{sec:BL_pre}
In this section, we describe how to construct the data matrices required for the spatial data-driven transient growth analysis of the boundary layer. There are four main steps: (i) form the state vector, (ii) take a spanwise Fourier transform, (iii) take a temporal Fourier transform, and (iv) form the final data matrix at each streamwise position. These steps are detailed below.  

The raw data from the JHTDB is of size $N_v \times N_x \times N_y \times N_z \times N_t$, where $N_v = 3$ is the number of velocity components, $N_x = 556$ is the number of retained streamwise positions, $N_y = 112$ is the number of wall-normal grid points, $N_z$ = 1024 is the number of spanwise grid points, and $N_t = 4701$ is the number of temporal snapshots. Since the applicable state vector ${\vectt{q}}$ for this problem consists of the three velocity components at each wall-normal grid point, we first reshape the data matrix such that the leading dimension contains the velocity components stacked together, giving a new data matrix of size $n \times N_x \times N_z \times N_t$, where $n = N_v N_y$ is the state dimension. Second, we take the spanwise discrete Fourier transform. Since each spanwise wavenumber $\beta$ is decoupled in linear analyses, this reduces the size of the data matrix for each spanwise wavenumber to $n \times N_x \times N_t$. 

Third, we perform the temporal Fourier transforms implied by \eqref{eqn:stateFT}. Since the flow is ergodic, we partition the single, long time series provided in the database into $m$ shorter blocks, each of which can be interpreted as a realization of the flow. This is analogous to Welch's method for computing power spectra and standard algorithms for spectral proper orthogonal decomposition \citep{Towne2018spectral}. Each block contains $N_f$ snapshots, which determines the frequency resolution, i.e., the frequency increment and total number of frequencies, after taking a discrete Fourier transform. 

Using blocks of finite length to approximate the infinite Fourier transform implied by \eqref{eqn:stateFT} can introduce unwanted effects. In particular, even if the data are entirely linear, \eqref{eqn:BL_state} is perfectly satisfied only in the limit of infinite intervals. Physically, this occurs because, if the interval is too short, the initial state ${\vectt{q}}_{0}$ does not have time to evolve and influence the downstream state ${\vectt{q}}_{x}$ beyond a certain point, leading to decorrelation between the initial and downstream data in each temporal block. This can be mitigated by shifting the temporal position of the blocks used to define the downstream pair of each initial condition. We do so by defining a convection velocity $U_c$ to determine how far into the future to shift a given downstream block to ensure that the data it contains is indeed the downstream response to the initial condition contained within the corresponding input block. Practically, we restrict the possible convection velocities to those values that result in blocks that align with the discrete data, i.e., $U_c = n_x \Delta x / n_t \Delta t$, where $n_x$ and $n_t$ are the numbers of spatial and temporal steps between the input and output blocks and $\Delta x$ and $\Delta t$ are the spatial and temporal spacing of the data. Details on determining the optimal $U_c$ can be found in \cite{kai2024data}. 

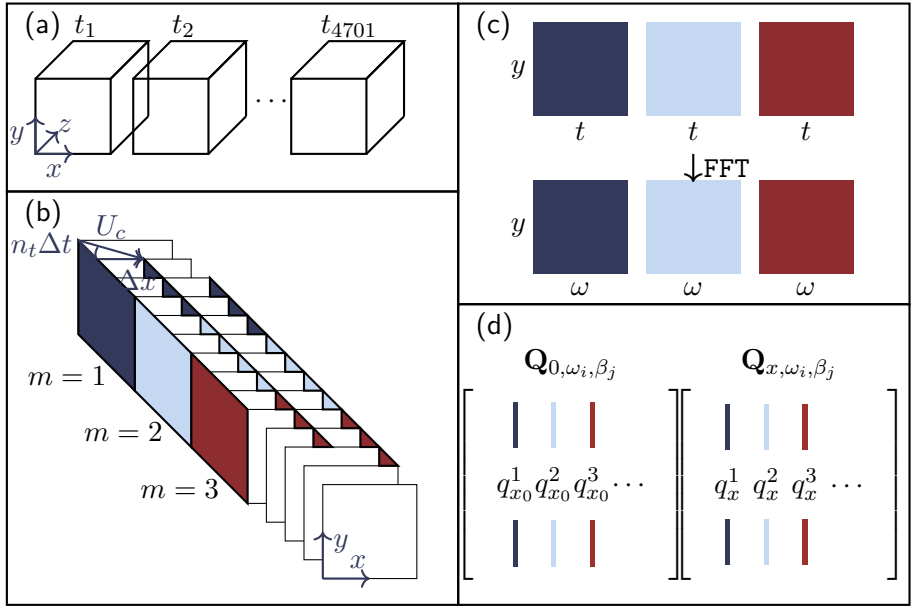
\begin{figure}[t]
    \centering
    \begin{tikzpicture}[font=\sffamily, scale=1, every node/.style={scale=1}]
\definecolor{color1}{RGB}{51,57,91}
\definecolor{color3}{RGB}{142,45,48}
\definecolor{color2}{RGB}{196,216,242}

  \draw[thick] (0,5.5) rectangle (6,8);
    \node[anchor=west] at (0.1,7.7) {(a)};
\foreach \x in {0,1.3,3.4} {
\draw[thick] (\x+0+0.33+0.05,0+6) -- ++(1,0) -- ++(0,1) -- ++(-1,0) -- cycle; 
  \draw[thick] (\x+0+0.33+0.05,1+6) -- ++(0.5,0.5) -- ++(1,0) -- ++(-0.5,-0.5); 
  \draw[thick] (\x+1+0.33+0.05,0+6) -- ++(0.5,0.5); 
  \draw[thick] (\x+1+0.33+0.05,1+6) -- ++(0.5,0.5); 
  \draw[thick] (\x+1.5+0.33+0.05,0.5+6) -- ++(0,1); 
}
  \node at (4.5-3.4-0.1+0.05,7.7) {\( t_{1} \)};
  \node at (4.5-2.1-0.1+0.05,7.7) {\( t_{2} \)};
  \node at (4.5+0.05,7.7) {\( t_{4701} \)};
  \node at (4.5-1+0.05, 7.7-1) {$\cdots$};
  
  \draw[->,color=color1,thick] (0.33+0.05,6) -- ++(0.5,0) node[midway, below] {$x$};
  \draw[->,color=color1,thick] (0.33+0.05,6) -- ++(0,0.5) node[midway, left] {$y$};
  \draw[->,color=color1,thick] (0.33+0.05,6) -- ++(0.3,0.3) node[midway, above right] {$z$};

  \begin{scope}[shift={(0,2.5)}]
    \draw[thick] (0,-2.5) rectangle (6,3);
    \node[anchor=west] at (0.1,2.7) {(b)};

    \foreach \j in {0,...,13} {
        \draw[fill=white, opacity=1] (0.2+0.75+\j*0.25, 1.625-\j*0.25-0.525) rectangle ++(1.25,1.25);
    }
\foreach \i in {0,...,1}{
    \foreach \j in {0,...,9} {
        \pgfmathparse{\j<3 ? "rgb,255:red,51; green,57; blue,91" : (\j>6 ? "rgb,255:red,142; green,45; blue,48" : "rgb,255:red,196; green,216; blue,242")}
        \edef\TriColor{\pgfmathresult}
        \draw[thick,fill=\TriColor, opacity=0.5]  (0.2+0.75+0.25+1.25/2+\j*0.25+\i*1.25/2+0.25*\i, 1.625-\j*0.25-0.25+1.25-\i*0.25-0.525) -- ++(0.25,-0.25) -- ++(-0.25,0) -- cycle;


    }
}
\foreach \j in {0,...,2}{
\pgfmathparse{\j==2 ? "rgb,255:red,51; green,57; blue,91" : (\j==0 ? "rgb,255:red,142; green,45; blue,48" : "rgb,255:red,196; green,216; blue,242")}
\edef\Reccolor{\pgfmathresult}
\draw[thick,fill=\Reccolor, opacity=0.5] (0.2+0.75+0.5+1-\j*0.75,0.75*\j-0.625+2-0.525) -- ++(0.75,-0.75) -- ++(0,-1.25) -- ++(-0.75,0.75) -- cycle;
}

    \node at (0.8,0.55) {$m=1$};
    \node at (0.8+0.75,0.55-0.75) {$m=2$};
    \node at (0.8+2*0.75,0.55-2*0.75) {$m=3$};
    \draw[->,color=color1,thick] (0.2+2.25-6*0.25,1.375+6*0.25-0.525) -- ++(1.25/2+0.25,-0.25) node[midway, above] {$U_c$};
    \draw[->,color=color1,thick] (0.2+2.25-6*0.25,1.375+6*0.25-0.525) -- ++(0.25,-0.25) node[near start, left] {$n_t \Delta t$};
    \draw[->,color=color1,thick] (0.2+2.25-6*0.25+0.25,1.375+6*0.25-0.25-0.525) -- ++(1.25/2,0) node[midway, below] {$n_x \Delta x$};
    \draw[->,color=color1,thick](0.2+0.75+13*0.25, 1.625-13*0.25-0.525)-- ++(1.25/2,0) node[near end, above] {$x$};
    \draw[->,color=color1,thick](0.2+0.75+13*0.25, 1.625-13*0.25-0.525)-- ++(0,1.25/2) node[near end, right] {$y$};
  \end{scope}

  \begin{scope}[shift={(6,5)}]
    \draw[thick] (0,-0.75-0.25) rectangle (6,3);
    \node[anchor=west] at (0.1,2.7) {(c)};
    \foreach \i/\col in {0/color1,1/color2,2/color3} {
      \fill[\col] (\i*1.5+1,1.5) rectangle ++(1.25,1.25);
      \fill[\col] (\i*1.5+1,-0.6) rectangle ++(1.25,1.25);
      \node at (\i*1.5+1+1.25/2,1.3) {$t$};
      \node at (\i*1.5+1+1.25/2,-0.8) {$\omega$};
    }
    
    \draw[->, thick] (1.5+1+1.25/2,1) -- (1.5+1+1.25/2,1.25/2) node[midway, right] {\texttt{FFT}};
    \node at (1-0.2,1.5+1.25/2) {$y$};
    \node at (1-0.2,-0.6+1.25/2) {$y$};

  \end{scope}

  \begin{scope}[shift={(6,1)}]

    \draw[thick] (0,-0.75-0.25) rectangle (6,3);
    \node[anchor=west] at (0.1,2.7) {(d)};
    \matrix[matrix of math nodes, nodes in empty cells, row sep=0cm, column sep=-0.25cm, left delimiter={[}, right delimiter={]}]at (1.6-0.1,0.8-0.2) (m1) {
     \color[RGB]{51,57,91}\rule[-2mm]{1.5pt}{6mm} & \color[rgb]{0.768627450980392,0.847058823529412,0.949019607843137}\rule[-2mm]{1.5pt}{6mm}&  \color[rgb]{0.611764705882353,0.192156862745098,0.160784313725490}\rule[-2mm]{1.5pt}{6mm} &\\
    q_{x_0}^1 & q_{x_0}^2 & q_{x_0}^3 & \cdots \\
    \color[RGB]{51,57,91}\rule[-2mm]{1.5pt}{6mm} & \color[rgb]{0.768627450980392,0.847058823529412,0.949019607843137}\rule[-2mm]{1.5pt}{6mm}&  \color[rgb]{0.611764705882353,0.192156862745098,0.160784313725490}\rule[-2mm]{1.5pt}{6mm} &\\
};

\matrix[matrix of math nodes, nodes in empty cells, row sep=0cm, column sep=-0.1cm, left delimiter={[}, right delimiter={]}]at (4.5-0.1,0.8-0.2) (m2) {
     \color[RGB]{51,57,91}\rule[-2mm]{1.5pt}{6mm} & \color[rgb]{0.768627450980392,0.847058823529412,0.949019607843137}\rule[-2mm]{1.5pt}{6mm}&  \color[rgb]{0.611764705882353,0.192156862745098,0.160784313725490}\rule[-2mm]{1.5pt}{6mm} &\\
    q_{x}^1 & q_{x}^2 & q_{x}^3 & \cdots \\
    \color[RGB]{51,57,91}\rule[-2mm]{1.5pt}{6mm} & \color[rgb]{0.768627450980392,0.847058823529412,0.949019607843137}\rule[-2mm]{1.5pt}{6mm}&  \color[rgb]{0.611764705882353,0.192156862745098,0.160784313725490}\rule[-2mm]{1.5pt}{6mm} &\\
};

\node at (1.5,2.2) {$\mathbf{Q}_{0,\omega_i,\beta_j}$};
\node at (4.4,2.2) {$\mathbf{Q}_{x,\omega_i,\beta_j}$};
  \end{scope}

\end{tikzpicture}
    \caption{The procedures for obtaining the input data matrices for our algorithm for the spatially evolving boundary layer. (a) The raw data is a time series of three-dimensional snapshots. (b) Take a spanwise discrete Fourier transform and select one wavenumber. The resulting time series is partitioned into a set of realizations; each color indicates a realization. Note that at downstream $x$-locations, the temporal interval for the same realization occurs later, with the delay determined by the convection velocity $U_c$ to maximize the correlation between the streamwise locations in the realization. (c) Take the temporal Fourier Transform for every realization. (d) Reorganizing the output from (c) such that states at the same $x$, $\omega$, and $\beta$ are in the same matrix, yielding the inputs to our method.}
    \label{fig:BL_pro}
\end{figure}

Finally, after taking the discrete Fourier transform of each block, the data at a given frequency of interest is extracted from each block and collected into the input and output matrices ${\matr{Q} }_{0}$, ${\matr{Q}}_{x} \in \mathbb{C}^{n \times m}$. The complete procedure described above is schematically depicted in Figure~\ref{fig:BL_pro}.   

\subsection{Boundary layer results}
\label{sec:BL_re}
Following \cite{andersson1999optimal} and \cite{luchini2000reynolds}, we focus on $\omega=0$ and consider a range of $\beta$ values. We set the regularization parameter to $\gamma_0 = 0.01$, based on the results for the Ginzburg-Landau case.

\begin{figure}[t]
    \centering
    \input{figure/plt_BL_Growth}
    \includegraphics{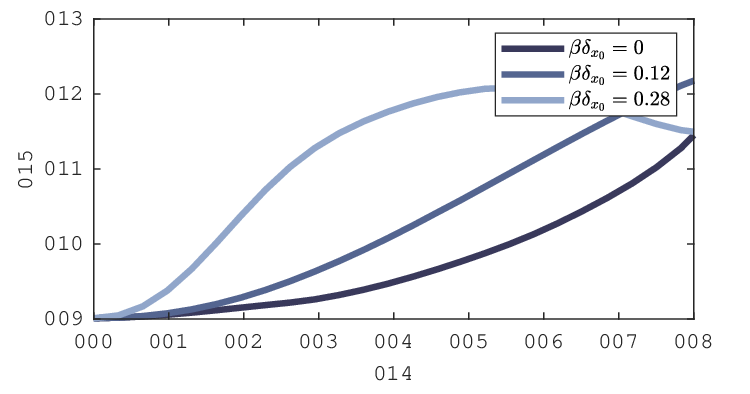}
    \caption{The spatial transient growth as a function of streamwise location returned by the data-driven method for three spanwise wavenumbers.}
    \label{fig:plt_BL_Growth}
\end{figure}

The growth over $x$ for three spanwise wavenumbers is depicted in Figure~\ref{fig:plt_BL_Growth}.
We observe that the growth for $\beta \delta_{x_0} = 0$ and $0.12$ is much slower than $\beta \delta_{x_0}=0.28$. The slower growth for zero $\beta$ is consistent with the slow growth of Tollmien-Schlichting waves known to be present at low $\beta$. All three curves are similar in magnitude to those given in \cite{andersson1999optimal} and \cite{luchini2000reynolds}. It bears mentioning that most curves, even the ones that decay at the end of the interval shown in Fig. 13, grow outside of this interval instead of decaying to zero as $Re_x \to \infty$. We believe this can be attributed to the breakdown of the linearity assumption as the flow transitions to turbulence. Also, we note that the magnitude of the growth is sensitive to $\gamma$. As we increase or decrease $\gamma$ by an order of magnitude, the growth also changes by an order of magnitude. Therefore, we rely on our tests with the Ginzburg-Landau equation, which suggests a possible value of $\gamma$, though the interpolation of the selection of $\gamma$ in different cases needs further investigation.

\begin{figure}[h!t]
    \centering
    \input{figure/plt_BL_Output}
    \includegraphics{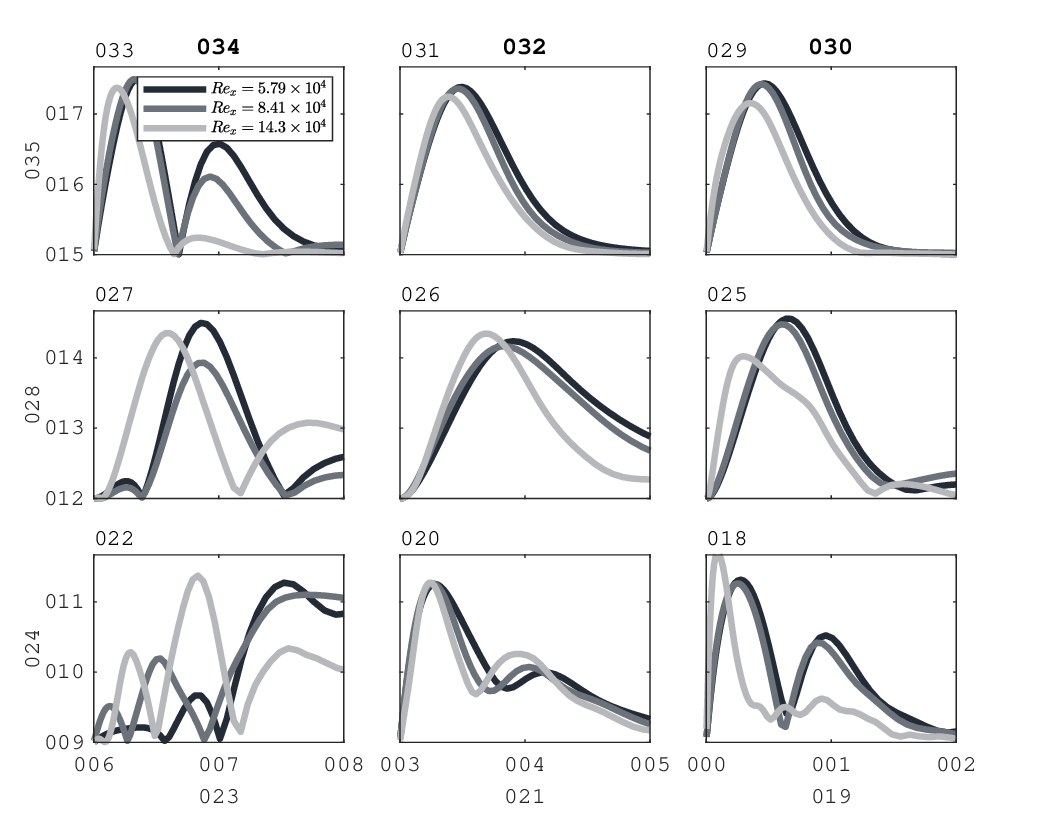}
    \caption{The output mode for the boundary layer for $\beta \delta_{x_0}$ of $0$, $0.12$, and $0.28$. Each row shows the output modes at different $x$ locations for a single component of the velocity field. A two-peak structure can be observed in (a), while in (b) and (c), there is only one peak.}
    \label{fig:plt_BL_Output}
\end{figure}

\begin{figure}[t]
    \centering
    \input{figure/plt_BL_Output_com}
    \includegraphics{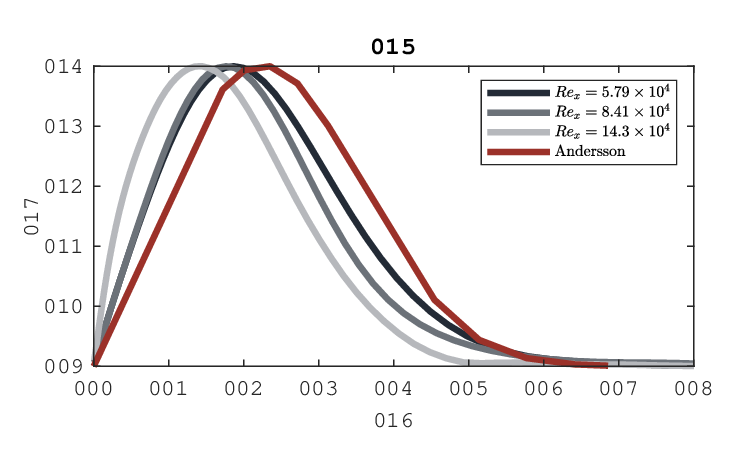}
    \caption{Comparison of the streamwise-velocity component of the output mode with the result in \cite{andersson1999optimal} for the boundary layer.}
    \label{fig:plt_BL_output_com}
\end{figure}

Figure~\ref{fig:plt_BL_Output} shows the optimal output mode for the three velocity components and the same three $\beta\delta_{x_0}$ values. Except for the $\beta=0$ results in the first row, we observe that the nonzero $\beta$ results are consistent over different $Re_x$ and $\beta$. The streamwise velocity $u$ and the wall-normal velocity $v$ display a one-peak structure, while the spanwise velocity $w$ has two peaks. At zero $\beta$, a two-peak structure is observed for the streamwise velocity disturbance. This structure resembles the Tollmien-Schlichting (T-S) modes described in \cite{schmid2002stability}, and it is well known that modal growth is strongest at $\beta = 0$.
\cite{andersson1999optimal} and \cite{luchini2000reynolds} have also included the output mode for the streamwise velocity disturbance. The comparison is shown in Figure~\ref{fig:plt_BL_output_com}. The curves for the streamwise velocity have a one-peak structure, which demonstrates good agreement with the results in ~\cite{andersson1999optimal} and \cite{luchini2000reynolds}, regardless of $x$ or $\beta$. As shown in the figure, the peak of the mode shifts to the right when $Re_x$ increases. The result from \cite{andersson1999optimal} is the limit of $Re_x\rightarrow \infty$, which is consistent with the trend in Figure~\ref{fig:plt_BL_output_com} that our curve shifts toward their result as $Re_x$ increases.

\begin{figure}[h!t]
    \centering
    \input{figure/plt_BL_GrowthVB}
    \includegraphics{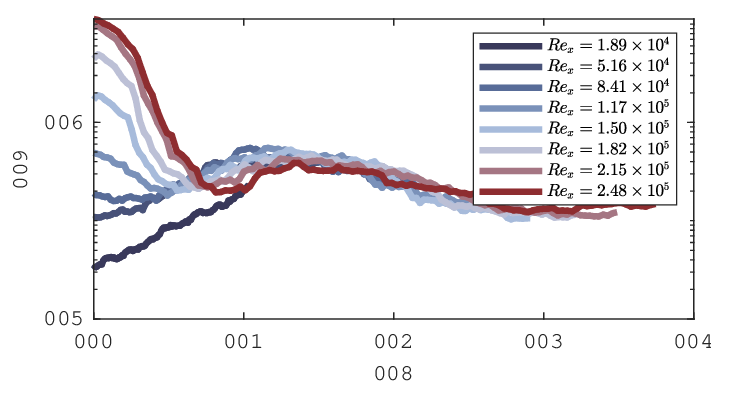}
    \caption{Growth as a function of the spanwise wavenumbers at different $Re_x$ and $\omega=0$. The curves are smoothed with $\mathtt{movmean}$ in MATLAB. The optimal growth for various $Re_x$ has a local peak at about $\beta\delta = 0.52$. The large growth at $\beta\delta = 0$ reflects the substantial modal growth possible at low $\beta$.}
    \label{fig:plt_BL_GrowthVB}
\end{figure}

\begin{figure}[h!t]
    \centering
    \input{figure/plt_gvw}
    \includegraphics{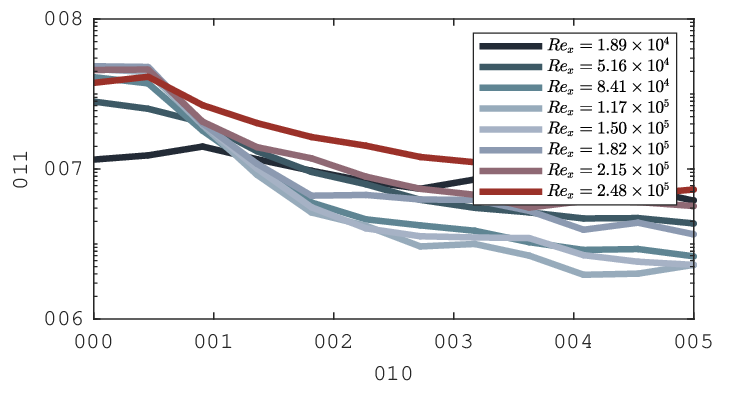}
    \caption{Growth as a function of the frequency for different $Re_x$. The maximum growth is observed at $\omega=0$.}
    \label{fig:plt_BL_GrowthvW}
\end{figure}

In Figure~\ref{fig:plt_BL_GrowthVB}, we show the growth normalized by $Re_x$ against $\beta \delta$; our results are consist with those of \cite{andersson1999optimal} and \cite{luchini2000reynolds} in several ways. These studies showed that $G^{opt}/Re_x$ approaches a constant in the limit $Re_x \to \infty$ and found that the asymptotic value of $G^{opt}/Re_x$ is maximized at $\beta \delta = 0.45$.
If we ignore the growth at $\beta\delta=0$ in Figure~\ref{fig:plt_BL_GrowthVB}, we can observe that there exists an optimal growth peak at $\beta\delta\approx 0.52$.
Additionally, the curves roughly overlap within the range of $Re_x$ in the figure.
What's more, the shape and magnitude of the growth are also similar to findings in~\cite{andersson1999optimal} and \cite{luchini2000reynolds}.
However, direct comparisons are not possible due to the difference in problem setup.
Finally, Figure~\ref{fig:plt_BL_GrowthvW} shows the growth over different frequencies, which has a maximum growth at $\omega=0$ for all values of $Re_x$. This is the same as stated in \cite{luchini2000reynolds} and justifies focusing on the result with $\omega=0$.


\section{Conclusions}\label{Sec:Conclusions}
We have developed a method that uses data to approximate the optimal transient growth curve and associated modes within the flow for both temporal and spatial stability problems. To estimate the growth and modes at a given time or streamwise location, the method requires a set of realizations of initial state vectors and the corresponding set of (temporally or spatially) evolved state vectors. Using these two sets -- and the principle of superposition -- we maximize the ratio of the energy of a linear combination of the final states and the energy of the same linear combination of initial states, over all linear combinations. We also regularize the method to mitigate the effect of noise, which may come in the form of measurement noise, external disturbances, or nonlinearity. 

The appeal of the proposed method is twofold. First, the linearized Navier-Stokes operator required in traditional transient growth calculations may be cumbersome to obtain from a large code. It also may be entirely unavailable, as in the case of an experiment. Conversely, the data required to apply the proposed method is readily available in many cases, making the method easier to use in many scenarios. Second, the proposed method scales much more favorably to large problems. While the traditional approach scales cubically in the spatial dimension due to the matrix exponential and SVD, the data-driven approach scales linearly in the spatial dimension and quadratically in the number of snapshots used – the same scaling as obtaining POD modes, which is not difficult for large problems. 

We validated the ability of the method to produce accurate estimates of these quantities for a linearized Ginzburg-Landau system, where the growth and modes can be computed analytically.
We also used the Ginzburg-Landau system to determine a reasonable choice for the regularization parameter in the method.
We then applied our method to the transitional boundary layer to identify spatial transient growth. Our prediction for the output mode of the streamwise velocity component aligns closely with operator-based results. From the output modes, we observed a clear distinction between the two-peak structure for modal growth and the one-peak structure for transient growth. We also identified a spanwise wavenumber that contributes significantly to transient energy growth.
The discrepancies between our results and the literature include differences in the growth over the streamwise location and the spanwise wavenumber that maximizes energy growth. We believe that much of the discrepancy is caused by the strong freestream disturbances triggering bypass transition present in the data, and that we would have seen better performance had we chosen a case in which linear growth played a more prominent role in the transition process. 

One particularly promising application of the method is studying hypersonic boundary layer transition. The transition process for high-speed flows is sensitive to numerous factors, and it can be challenging to include all of the relevant physics within a linear stability solver that could be used to conduct an operator-based transient growth analysis. In contrast, excellent data exists for these flows that could be used as input to our data-driven transient-growth algorithm.

\begin{appendices}

\section{Least-squares approximation method}
\label{sec:LSAM}
Here, we show another formulation of the data-driven method inspired by dynamic mode decomposition (DMD) \citep{SCHMID2010} that is equivalent to the formulation of the method presented in the main text in the case of no regularization. In addition to offering another perspective on the method that may be more intuitive to some readers, this formulation suggests alternative regularization strategies.

The assumption of linearity guarantees that there is a matrix $\matr{M}$ such that
\begin{equation}
    \matr{Q}_t=\matr{MQ}_0.
    \label{eqn:dmd_pseudo}
\end{equation}
For illustration purposes, we show the unweighted case here, i.e., $\matr{W}=\matr{I}$.
Analogous to DMD, we approximate $\matr{M}$ as
\begin{equation} \label{eq:app:DMD}
    \hat{\matr{M}} = \matr{Q}_t\matr{Q}_0^+ \approx \matr{M},
\end{equation}
where $\matr{Q}_0^+=\left(\matr{Q}_0^*\matr{Q}_0\right)^{-1}\matr{Q}_0^*$ (as we have assumed $Q_0$ has full column rank).
The optimal growth is then approximated using this estimation of the evolution operator as 
\begin{equation}
G_{LSA}^{opt}(t)=\sigma_1^2\left(\hat{\matr{M}}\right), \label{eqn:glsa}
\end{equation}
The similarity between this method and DMD lies in the way they both approximate the linear operator relating two data matrices; the differences between the two methods are that i) the corresponding columns in the data matrices in DMD are a single time step apart, whereas here they are separated by the time between the input and output disturbance, and ii) DMD takes an eigendecomposition of the resulting linear operator and we take an SVD.
\\

This formulation is equivalent to that presented in the main text in the unregularized case, and this can be shown as follows. Leveraging the assumption that $\matr{Q}_{0}$ has full column rank, we first take the Cholesky decomposition
\begin{equation}
    \matr{B}_{UW}^*\matr{B}_{UW} = \matr{Q}_0^*\matr{Q}_0, \label{eqn:cholesky_TLS}
\end{equation}
where the subscript is a reminder that we are in the unweighted case. Now, \eqref{eqn:glsa} may be rewritten as 
\begin{equation}
    G_{LSA}^{opt}(t) = \sigma_1^2 \left( \matr{Q}_t \matr{B}_{UW}^{-1} \matr{B}_{UW}^{*-1} \matr{Q}_{0}^* \right) \text{.}
\end{equation}
The square of the largest singular value of any matrix is the largest eigenvalue of the same matrix multiplied by its Hermitian conjugate. Using this fact, we have
\begin{equation}
    G_{LSA}^{opt}(t) = \lambda_1 \left( \matr{Q}_t \matr{B}_{UW}^{-1} \matr{B}_{UW}^{*-1} \matr{Q}_{0}^*  \matr{Q}_{0} \matr{B}_{UW}^* \matr{B}_{UW}^{*-1} \matr{Q}_t^* \right) \text{.}
\end{equation}
Using \eqref{eqn:cholesky_TLS}, this simplifies to
\begin{equation}
    G_{LSA}^{opt}(t) = \lambda_1 \left( \matr{Q}_t \matr{B}_{UW}^{-1} \matr{B}_{UW}^{*-1} \matr{Q}_t^* \right) \text{.}
\end{equation}
Finally, using the correspondence between singular values and eigenvalues, we have
\begin{equation} \label{eqn:DMD_noweight}
    G_{LSA}^{opt}(t) = \sigma_1^2 \left( \matr{Q}_t \matr{B}_{UW}^{-1}  \right) \text{.}
\end{equation}
Equation~\eqref{eqn:DMD_noweight} is equivalent to Equation \eqref{eqn:CFM_re} in the case of no weight matrix, i.e., $\matr{L} = \matr{I}$. It can also be shown that
\begin{equation}
\begin{aligned}
    \vectt{q}^{opt}_{0} &= \matr{Q}_0 \vectt{\psi}_{UW,1}/\left\|\matr{Q}_0 \vectt{\psi}_{UW,1}\right\|,\\
    \vectt{q}^{opt}_{t} &= \matr{Q}_t \vectt{\psi}_{UW,1}/\left\|\matr{Q}_t \vectt{\psi}_{UW,1}\right\|. \label{eqn:IO_2_TLS}
\end{aligned}
\end{equation}
where $\psi_{UW,1}=\matr{B}^{-1}_{UW}v_{UW,1}$, and $v_{UW,1}$ is the first column of the right singular vector of $ \matr{Q}_t \matr{B}_{UW}^{-1}$. Additionally, $\vectt{q}_t^{opt}$ is equivalent to the left singular vector of $ \matr{Q}_t \matr{B}_{UW}^{-1}$, as shown in~\cite{kai2024data}.
\\

The DMD-based approach in the weighted case approximates $\matr{M}$ by first transforming to a coordinate system where the $2-$norm is equivalent to the $\matr{W}$ norm in the original coordinates, then approximating $\matr{M}$ with a pseudoinverse in these coordinates. Mathematically, 
\begin{equation}
    \hat{\matr{M}} = \matr{Y}_t\matr{Y}_0^+
\end{equation}
where $\matr{Y}_t = \matr{LQ}_t$ and $\matr{Y}_0 = \matr{LQ}_0$ are the matrices of weighted states. The growth is again given by the SVD of $\hat{\matr{M}}$, and this may be shown to be equivalent to
\begin{equation}
   G_{LSA}^{opt}(t) = \sigma_1^2 \left( \matr{L} \matr{Q}_t \matr{B}_{UW}^{-1}  \right) \text{,}
\end{equation}
which is the same as the growth in the unregularized, weighted case in the main text. The modes can also be shown to be the same. 
\\

\begin{figure}[t]
    \centering
    \include{figure/plt_GL_m_err_TLS}
    \includegraphics{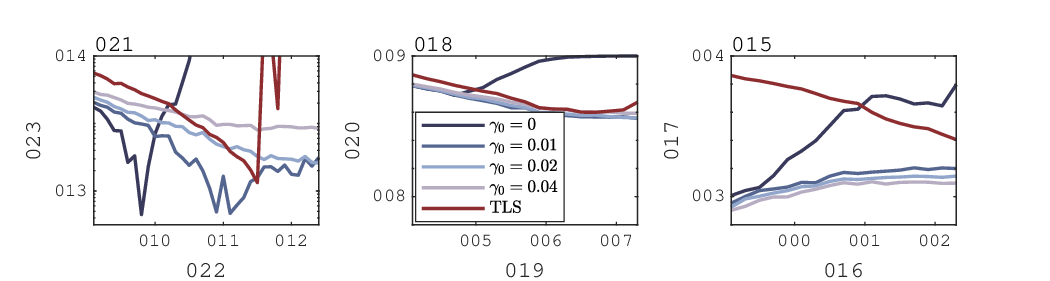}
    \caption{Same as Figure~\ref{fig:plt_GL_m_err}, with the additional red curve representing the error of TLS.}
    \label{fig:plt_GL_m_err_TLS}
\end{figure}

\begin{figure}[h!t]
    \centering
    \input{figure/plt6_compaddtls_tls}
    \includegraphics{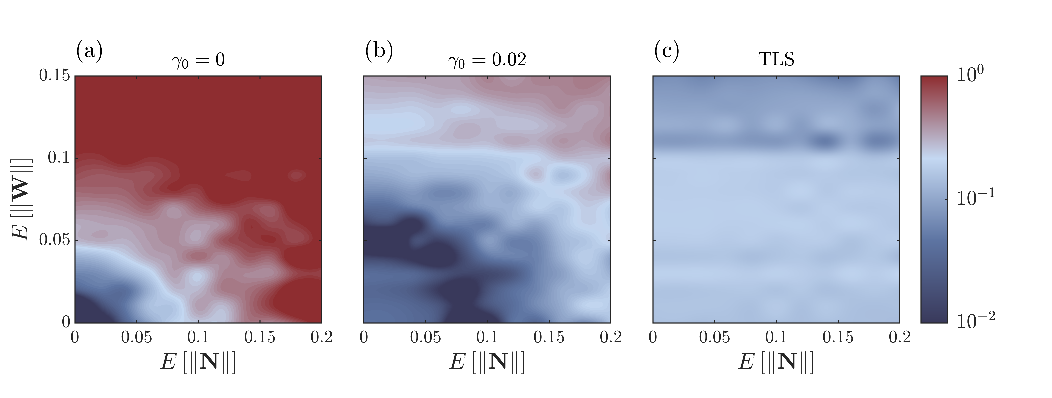}
    \caption{Peak growth error $\epsilon_G$ for the Ginzburg-Landau system as a function of measurement and process noise levels: (a) unregularized; (b) the regularization from Section~\ref{sec:Regularization} with $\gamma_0 = 0.02$; (c) TLS regularization.}
    \label{fig:noiseTLS}
\end{figure}

The potential advantage of this equivalence and the DMD-based approach is that we can now regularize it with methods originally designed for DMD. Here, we choose the total least squares (TLS) method from~\cite{Tu2014}. As shown in Figure~\ref{fig:plt_GL_m_err_TLS} and~\ref{fig:noiseTLS}, this method generally works for the test case with the Ginzburg-Landau equation.
Figure~\ref{fig:plt_GL_m_err_TLS} is equivalent to Figure~\ref{fig:plt_GL_m_err} in the main text but with the addition of the results for TLS regularization. Note that the $\gamma_0=0$ case is equivalent to the DMD-based method without regularization. Compared with the regularization method in previous sections, the total least squares method depends only on the number of realizations, $m$, but produces a less accurate result in general. As shown in Figure~\ref{fig:plt_GL_m_err_TLS}, more realizations allow the total least squares method to get better estimations of the modes. However, the growth estimated by the total least squares is very sensitive to noise when $m$ is large. Additionally, the TLS regularization produces less accurate output modes than the method in the main text for all $m$ values.
Figure~\ref{fig:noiseTLS} plots the peak growth error, $\epsilon_G$, under different measurement and process noise with $m=110$. In this figure, we can observe that the TLS method can produce reasonable results over the range of noise. Although the $\epsilon_G$ for TLS is higher than the regularization proposed in Section~\ref{sec:Regularization}, it still outperforms the unregularized method.

\begin{figure}[h!t]
    \centering
    \input{figure/plt_BL_Output_TLS}
    \includegraphics{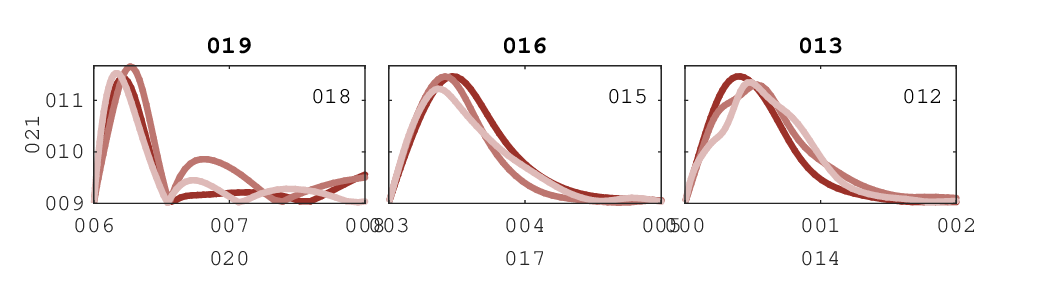}
    \caption{The output mode for the boundary layer at various streamwise locations and for three spanwise wavenumbers computed using the TLS method.}
    \label{fig:plt_BL_Output_TLS}
\end{figure}

For the boundary layer data, the TLS regularization can only capture the optimal output modes as shown in Figure~\ref{fig:plt_BL_Output_TLS}.
\begin{figure}[h!t]
    \centering
    \input{figure/plt_BL_Growth_TLS}
    \includegraphics{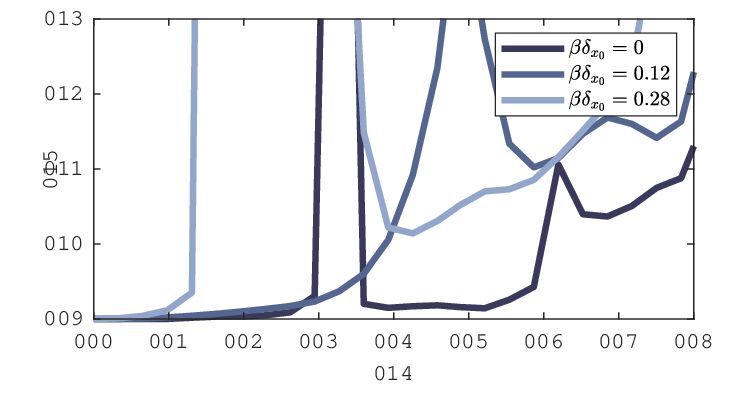}
    \caption{The spatial transient growth for the boundary layer as a function of streamwise location for three spanwise wavenumbers computed using the TLS method.}
    \label{fig:plt_BL_Growth_TLS}
\end{figure}
The output mode results are consistent with the other method. However, the growth result in Figure~\ref{fig:plt_BL_Growth_TLS} is obviously dominated by noise.
The cause of this is not entirely clear; one potential explanation could be the insufficient number of realizations, $m$. Since the TLS method involves initial projection into a low-dimensional POD space, this step could significantly reduce the accuracy of the result. In addition, treating the nonlinearities as noise may exceed the capabilities of the TLS method, since its development in \cite{Tu2014} aims to suppress process noise without considering nonlinearities.

\end{appendices}

\bibliographystyle{jfm}
\bibliography{main}

\end{document}